\documentclass[aps,reprint,twocolumn,superscriptaddress,10pt]{revtex4-1}
\usepackage{siunitx}
\usepackage{amsfonts}
\usepackage{amsmath}
\usepackage{amssymb}
\usepackage{bm}
\usepackage{graphicx}
\usepackage{dcolumn}
\usepackage{mciteplus}
\usepackage{euscript}
\usepackage{multirow}
\usepackage{color}
\usepackage{textcomp}
\usepackage{gensymb}
\usepackage{dblfloatfix}
\usepackage{float}
\DeclareMathOperator{\Tr}{Tr}

\newcommand{\er}{E^\text{so}}
\newcommand{\vso}{V}
\newcommand{\ef}{\epsilon_\text{F}}
\newcommand{\gfull}{\widetilde{G}}
\newcommand{\gb}{G}

\newcommand{\bfR}{{\bf R}}

\newcommand{\bfk}{{\bf k}}

\newcommand{\redbf}[1]{}

\newcommand{\memome}[1]{}
\newcommand{\req}[1]{Eq.~(\ref{#1})}

\newcommand{\rfig}[1]{Fig.~\ref{#1}}
\newcommand{\rFig}[1]{Figure~\ref{#1}}
\newcommand{\rtbl}[1]{Table~\ref{#1}}
\newcommand{\mw}[1]{\multirow{2}{*}{#1}}

\newcommand{\ud}{\mathrm{d}}
\newcommand{\uvecn}{{\bm{\hat{\textnormal{\bf n}}}}}

\newcommand\abinitio{\emph{ab initio}}

\begin{document}
%% \title{Two-ion magnetocrystalline anisotropy using maximally localized Wannier functions }
\title{Intersublattice magnetocrystalline anisotropy using a realistic tight-binding method based on maximally localized Wannier functions}

\author{Liqin Ke}
\email{liqinke@ameslab.gov}
\affiliation{Ames Laboratory, U.S.~Department of Energy, Ames, Iowa 50011}

\begin{abstract}

Using a realistic tight-binding Hamiltonian based on maximally localized Wannier functions, we investigate the two-ion magnetocrystalline anisotropy energy (MAE) in $L1_0$ transition metal compounds.
MAE contributions from throughout the Brillouin zone are obtained using magnetic force theorem calculations with and without perturbation theory.
The results from both methods agree with each other, and both reflect features of the Fermi surface.
The intrasublattice and intersublattice contributions to MAE are evaluated using a Green’s function method.
We find that the sign of the intersublattice contribution varies among compounds, and that its amplitude may be significant, suggesting MAE can not be resolved accurately in a single-ion manner.
The results are further validated by scaling spin-orbit-coupling strength in density functional theory.
Overall, this realistic tight-binding method provides an effective approach to evaluate and analyze MAE while retaining the accuracy of corresponding first-principles methods.

\end{abstract}

\date{\today}
\maketitle

\section{Introduction}

\redbf{Importance of SOC and MAE} Magnetocrystalline anisotropy (MA) arises from the interplay between spin-orbit coupling (SOC) and crystal-field effects and is one of the most fundamental intrinsic magnetic properties~\cite{van-vleck1937pr}.
Materials with high MA have been used in many applications including permanent magnets~\cite{mccallum2014armr} and magnetic recording media. One of the latest examples is the recently realized magnetic two-dimensional (2D) van der Waals (vdW) class of materials, in which MA is required to stabilize the long-range magnetic ordering down to atomically thin dimensions. These materials can be exploited as platforms for true 2D magnetism and for innovative applications such as energy-efficient, ultra-compact, spin-based electronics.
In general, it is of great interest to evaluate and resolve MA, and to unravel the underlying mechanisms in a given system.
Ultimately, such understanding will guide the control and tuning of MA, accelerating the development of new materials and their applications~\cite{rau2014s}.

Density functional theory (DFT) has proven to be a valuable tool to investigate and predict MA energy (MAE) in various systems.
The relativistic effects of valence electrons are often treated using various approximations to reduce computational complexity and cost.
Instead of directly solving the four-component Dirac equation self-consistently, one usually treats SOC as perturbation and starts first with the two-component scalar-relativistic (SR) Hamiltonian~\cite{koelling1977jpcs}, omitting SOC but including all other relativistic effects such as mass-velocity and Darwin terms. SOC can be added directly into the SR Hamiltonian or included in a subsequent step using the basis (often a subset of it) of SR wave functions (second variation)~\cite{li1990prb,shick1997prb}. Because the charge- and spin-density variations caused by SOC vanish to first order in the SOC strength~\cite{wang1996jmmm}, the magnetic force theorem (MFT)~\cite{mackintosh1980book,weinert1985prb,daalderop1990prb} is often applied on top of the second-variation method to calculate MAE as the difference of one-electron band energies.
Finally, perturbation theory (PT)~\cite{yosida1965ptp,abate1965pr,takayama1976prb,bruno1989prb,cinal1994prb,van-der-laan1998jpcm,ke2015prb,jesche2015prb} is also often used with the MFT to compute and analyze the change of band energies due to SOC. Overall, depending on the system size and approximations used, MAE computation can be quite demanding, because of the enlarged dimension of Hamiltonian and the reduced symmetry due to SOC, and the denser $k$ mesh needed for high accuracy.

\redbf{MAE computation} Empirical or semi-realistic tight-binding (TB) methods~\cite{bruno1989prb} were widely used to study MAE long before MAE became accessible to more sophisticated DFT methods~\cite{daalderop1990prb}.
Pioneering work~\cite{bruno1989prb,cinal1994prb} using TB provided a fundamental understanding of MAE in various systems.
However, empirical TB Hamiltonians are generally hard to parametrize and often have insufficient accuracy to describe band structure, usually limiting TB to obtaining qualitative results in systems with large MAE.
The recently developed maximally localized Wannier functions (MLWFs) method~\cite{marzari1997prb,marzari2012rmp,souza2001prb} has been widely used to effectively construct TB Hamiltonians to compute many properties such as Fermi surfaces, Berry curvature, and transport.
With a smaller basis, it can describe an isolated set of bands and/or entangled bands in a given energy window.
Indeed, this method is also very suitable for MAE calculations, considering that MAE is, after all, a ground-state quantity, determined by the occupied states.
However, due to the minuteness of MAE, it is not clear how accurately the realistic TB can describe MAE in systems such as transition metal bulk compounds.
Here, we demonstrate that the realistic TB framework based on the MLWFs method can produce accurate MAE in comparison to DFT, thereby providing an efficient framework to compute and analyze MAE.

Various decomposition schemes have been used to resolve MAE into $k$ space, atomic sites, orbital, and spin channels, providing insight and guidance on tuning MAE.
The MFT enables resolution of MAE into individual bands on each $k$ point in reciprocal space, allowing a band-structure-origin analysis of MAE.
Site-resolved MAE is often calculated using methods such as evaluation of the on-site SOC energy~\cite{antropov2014ssc}, second-order PT, scaling the SOC strength, or others~\cite{wang1996prb,aberg2015tech-osma}.
For example, $R_2$Co$_{17}$ (with $R$ = Y or Ce) compounds have very small MAE and {\abinitio} analysis found that one particular Co sublattice, the so-called dumbbell sites, has a large negative contribution to uniaxial anisotropy~\cite{ke2016prbA}.
Thus, proper substituents that preferentially occupy the dumbbell sites and eliminate the negative contributions will significantly improve the uniaxial anisotropy, as observed in experiments.
Another interesting example is the oscillating MAE behavior in diluted nitridometalates Li$_{2}$[(Li$_{1-x}T_{x}$)N], with $T$ = Mn, Fe, Co, or Ni.
The MAE can be solely attributed to individual $T$ atoms and the band-filling effect on MAE can be quantitatively described in a single-ion MAE model~\cite{ke2015prb,jesche2015prb}.

Site-resolved MAE values, together with exchange parameters, can also be used as inputs for subsequent large-scale atomic spin simulations to calculate temperature-dependent magnetic properties.
When interfacing $\abinitio$ results with atomic spin simulation, MAE is usually treated in the single-ion model in the atomic spin Hamiltonian.
However, it has been found that a two-ion anisotropy model is needed to properly describe properties such as temperature-dependent MAE in some systems~\cite{mryasov2005eel}.
Thus, it is of great interest to resolve MAE into sites and, in particular, beyond the single-ion model.

In this work, we investigate two-ion MAE in various $L1_0$ compounds, one of the most widely studied systems, with MAE values ranging from several tens of \si{\mu eV} to a couple of \si{\meV} per formula unit~\cite{daalderop1991prb,solovyev1995prb,kota2014jpsj}.
Our approach is based on second-order PT using a Green's function method implemented within the realistic TB framework.
We demonstrate that this approach achieves accuracy similar to DFT and provides a highly efficient means to compute and analyze the two-ion anisotropy in transition metal systems.

\section{Computational Details}
\label{sec:2}

We first construct the real-space scalar-relativistic TB Hamiltonian using the MLWFs method.
The corresponding Green's function is also constructed for use in the PT approach.
MAE is calculated using the MFT in TB, with and without PT, referred to hereafter as PT and MFT, respectively, for simplicity.
DFT methods, including both \textsc{vasp}~\cite{kresse1993prb,kresse1996prb} and an all-electron full-potential LMTO (FP-LMTO) method~\cite{methfessel2000coll-c3fl}, are used to calculate MAE and compare with TB.
To compare with the two-ion MAE values obtained using PT in TB, we also calculate intersublattice MAE contribution by scaling the SOC strength in \textsc{vasp}.
All DFT calculations are carried out within the generalized gradient approximation (GGA) using the functional of Perdew, Burke, and Ernzerhof (PBE)~\cite{perdew1996prl} unless local density approximation (LDA)~\cite{von-barth1972jpcs} is specified.

\subsection{TB Hamiltonian and SOC}

The MLWFs are constructed through a postprocessing procedure~\cite{marzari1997prb,souza2001prb,marzari2012rmp} using the output of a self-consistent scalar-relativistic \textsc{vasp} calculation.
For each $L1_0$ compound, 18 MLWFs corresponding to $s$-, $p$-, and $d$-type orbitals for each of the two atoms in the unit cell were generated using \textsc{wannier90}~\cite{mostofi2014cpc}.
The spread functional for entangled energy bands is minimized by a two-step procedure~\cite{souza2001prb}.
An outside energy window with a larger number of bands was selected to ensure good description of the band structure of the ``frozen'' inner energy window, spanning from the bottom of the valence band to a few eV above the Fermi level.
A real-space Hamiltonian $H(\bfR)$ with dimensions 18$\times$18 is constructed to accurately represent the band structures in this specified ``frozen'' energy window.
Then, $H({\bf k})$ is obtained by Fourier transformation.
The energy bands are recalculated within TB to ensure that DFT bands can be accurately reproduced before further MAE calculations.

The TB Hamiltonian is represented in a basis of orthonormalized atomic functions $|i,l,m,\sigma\rangle$, where $i$ labels atomic sites, $l, m$ angular and magnetic quantum numbers (in cubic harmonics), and $\sigma$ the spin.
The SOC part of the Hamiltonian, which can be directly added into $H$ or included using PT, can be written as

\begin{equation}
V_\text{so}=\xi~\mathbf{L}{\cdot}\mathbf{S}=\frac{\hbar^2}{2M^2c^2}\frac{1}{r}\frac{\ud
  V}{\ud r}{\bf L}\cdot{\bf S},
\end{equation}
where ${\bf L}\cdot{\bf S}$ depends explicitly on the direction of spin quantization axis (details can be found in Appendix~\ref{seca:soc}).
The radial part of $V_\text{so}$, the SOC constants $\xi_{i,l}^{\sigma\sigma'}$, are calculated using FP-LMTO.
For simplicity, we ignore the energy and spin dependence of $\xi$.
Furthermore, the occupation numbers of the Pt-$p$ orbitals, which have a large SOC constant, are overestimated in TB in comparison to DFT.
We renormalize the Pt-$p$ occupation numbers based on the DFT value when adding SOC into the Hamiltonian.

\subsection{MAE}
Turning on SOC lowers the system energy.
Here, we refer to this energy change as the SOC energy $\er$, which depends on the spin direction.
For uniaxial geometry, MAE can be defined as $K=\er_{110}-\er_{001}$, with $\er_{110}$ and $\er_{001}$ indicating the SOC energies along the spin directions $[110]$ and $[001]$, respectively.
We use $[110]$ as the reference direction for the basal plane.
A positive $K$ value indicates the system has uniaxial anisotropy with the easiest spin direction being out of plane.

\subsubsection{Magnetic force theorem}
After SOC is added into the TB Hamiltonian, tetrahedron integration with Bl\"ochl correction~\cite{blochl1994prb} is used to determine the Fermi level $\epsilon_\text{F}$, band weights, and band sums.
MAE is calculated as

\begin{equation}
K=\sum_{k,b}\left(
\epsilon^{110}_{k,b}f^{110}_{k,b}-\epsilon^{001}_{k,b}f^{001}_{k,b}\right),
\label{eqn_k_esum}%
\end{equation}
where, $b$ is the band index, $k$ refers to the wave vector in the first Brillouin zone (BZ), and $f_{k,b}$ is the corresponding band occupancy.
In order to resolve the MAE into $k$ space, \req{eqn_k_esum} needs to be modified.
A grand-canonical ensemble version~\cite{barreteau2016crp} is used:

\begin{equation}
  K=\sum_{k,b}\left( (\epsilon^{110}_{k,b}-\epsilon^{0}_\text{F})
  f_{k,b}^{110} -(\epsilon^{001}_{k,b}-\epsilon^{0}_\text{F})
  f_{k,b}^{001} \right),
\label{eqn_k_esum_ef0}%
\end{equation}
where $\epsilon^{0}_\text{F}$ is the Fermi level calculated without SOC.
The total MAE value does not depend on the reference energy because the total number of valence electrons $\sum f_{k,b}$ is conserved when spin is along different directions; however, the $k$-resolved MAE does depend on the choice of reference energy~\cite{subkow2009prb,aberg2015tech-osma}.
We use \req{eqn_k_esum_ef0} to properly decompose MAE because $k$ resolution of MAE based on \req{eqn_k_esum} will only reflect the change of band occupancy $\Delta f_{k,i}$.
The \req{eqn_k_esum} result will be dominated by the relaxation effect of the Fermi surface due to turning on SOC with magnetization along various directions, and MAE contributions will only be significant near the Fermi surface.

\subsubsection{Perturbation theory}
Using second-order PT, we can express orbital moment, SOC energy, and their anisotropies in terms of the susceptibility~\cite{ke2015prb} calculated using the unperturbed band structure.
The SOC energy $\er$ due to the spin-orbit interaction $V_\text{so}$ can be written as

\begin{equation}
\er=-\frac{1}{2\pi}%
\operatorname{Im}{\int_{-\infty}^{\epsilon_{F}}{\ud\epsilon} \Tr\left[\gfull(\epsilon) V_\text{so}\right]},%
\label{eqn_k_full}%
\end{equation}
where $\gfull(\epsilon)$, the full Green's function, includes SOC and can be constructed from the non-perturbed Green's function ${G}(\epsilon)$.
Using second-order PT (here we consider only systems with a uniaxial geometry), the SOC energy can be written as

\begin{equation}
\begin{aligned} 
\label{eqn_k2_ij}%
\er &= -\frac{1}{2\pi}\operatorname{Im}{\int_{-\infty}^{\ef}\ud\epsilon \Tr\left[G(\epsilon)\vso G (\epsilon)\vso\right]}, \\  %
\end{aligned}
\end{equation}
where the Green's function is constructed using
\begin{equation}
G(\epsilon)=\left(\epsilon-H\right)^{-1}. \\
\end{equation}

A complex contour integration on an elliptical path~\cite{zeller1982ssc} is used for the integration.
By exploiting the fact that $\gb$ is spin diagonal and $V$ is ($i,l$) diagonal, the SOC energy can be written as

\begin{equation}
  \er(\uvecn) = \sum_{il;jl',\sigma\sigma'}E_{il,jl'}^{\sigma,\sigma'}(\uvecn),
\end{equation}
where $E_{il,jl'}^{\sigma,\sigma'}(\uvecn)$ is the contribution from the sublattice-orbital-spin pair ($il\sigma,jl'\sigma'$).
We have

\begin{equation}
  E_{il,jl'}^{\sigma\sigma'}(\uvecn) =
  -\frac{1}{2\pi}\operatorname{Im}{\int_{-\infty}^{\ef}\ud\epsilon\int\ud\bfk\Tr
    E_{il,jl'}^{\sigma\sigma'}(\bfk,\epsilon;\uvecn)},
\end{equation}
and 
\begin{equation}
  E_{il,jl'}^{\sigma\sigma'}(\bfk,\epsilon;\uvecn) = G_{il,jl'}^{\sigma}(\bfk,\epsilon)\vso_{jl'}^{\sigma\sigma'}(\uvecn)G_{jl',il}^{\sigma'}(\bfk,\epsilon)\vso_{il}^{\sigma^\prime\sigma}(\uvecn).
\end{equation}

$\vso=\xi_{i,l}^{\sigma\sigma'}(\epsilon)\mathbf{L}{\cdot}\mathbf{S}(\uvecn)$ couples states within the same $l$ channel at the same site.
Here, for simplicity, we treat SOC strength as a constant $\xi_{il}$ for each $l$ channel at site $i$, ignoring its energy and spin dependence.
$\mathbf{L}{\cdot}\mathbf{S}$ can be written as a function of magnetization direction $\uvecn$.
The MAE can be written as

\begin{equation}
K=\sum_{ij,\sigma\sigma'}
K_{ij}^{\sigma\sigma'}=\sum_{ij,\sigma\sigma'}
E_{ij}^{\sigma\sigma'}(\uvecn_{110})-E_{ij}^{\sigma\sigma'}(\uvecn_{001}).
\label{eq:k1}
\end{equation}

We define the isotropic and anisotropic parts of $V(\uvecn)$ as $U$ and $A$, respectively:

\begin{align}
  \begin{split}
    2U &= V(\uvecn_{110})+V(\uvecn_{001}), \\ 
    2A &= V(\uvecn_{110})-V(\uvecn_{001}).
  \end{split}
\end{align}
Then, \req{eq:k1} can also be written as
\begin{equation}
  K =\sum_{ij,\sigma\sigma'} \widetilde{K}_{ij}^{\sigma\sigma'}
  \label{eq:kprime}
\end{equation}
with
\begin{equation}
\begin{aligned}
  \widetilde{K}_{ij}^{\sigma\sigma'} = &
  -\frac{2}{\pi}\operatorname{Im}\int_{-\infty}^{\ef}\ud\epsilon\int\ud\bfk \\ 
&   \Tr \left[
    G_{ij}^{\sigma}(\bfk,\epsilon)U_{j}^{\sigma\sigma'}G_{ji}^{\sigma'}(\bfk,\epsilon)A_{i}^{\sigma^\prime\sigma}\right]. 
\end{aligned}
\label{eq:k2}
\end{equation}

Here, we have
$\widetilde{K}_{ij}^{\sigma\sigma'}={K}_{ij}^{\sigma\sigma'}$ when
$\sigma=\sigma'$. Unlike
${K}_{ij}^{\uparrow\downarrow}={K}_{ij}^{\downarrow\uparrow}$, we
have
$\widetilde{K}_{ij}^{\uparrow\downarrow}\neq\widetilde{K}_{ij}^{\downarrow\uparrow}$,
however,
\begin{equation}
\widetilde{K}_{ij}^{\uparrow\downarrow}+\widetilde{K}_{ij}^{\downarrow\uparrow}
= {K}_{ij}^{\uparrow\downarrow}+{K}_{ij}^{\downarrow\uparrow}.
\end{equation}

According to \req{eq:k2}, the strength of the intersublattice MAE contribution $K_{ij}$ depends on the SOC strength of both sublattices $\xi_i$ and $\xi_j$ (contained in $V_\text{so}$ or $U$ and $A$) and on the intersublattice Green's function $G_{ij}$.
The element types of the sublattices determine the SOC strength $\xi$ while $G_{ij}$ is relevant to the hopping or hybridization between two sublattices and depends on the detail of electronic structure.
If there is no hybridization between the two sublattices or they are coupled only through $s$ orbitals (i.e., the angular parts of $U$ and $A$ vanish) elements, the corresponding intersublattice contribution becomes negligible.

\subsubsection{Scaling SOC strength}
Instead of using \req{eq:k2}, the intrasublattice and intersublattice MAE contributions can also be obtained by scaling the SOC strength $\xi_i$ on each site $i$ by a factor $\lambda_i$, and by fitting the MAE as a function of scaling vector $\bm{\lambda}=[\lambda_1, \lambda_2, ..., \lambda_n]^\intercal$:

\begin{equation}
V_\text{so}(\bm{\lambda})=\sum_{i} \lambda_i \xi_i ~\mathbf{L}{\cdot}\mathbf{S},
\end{equation}
\begin{equation}
  \begin{aligned}
K(\bm{\lambda})&=\sum_{ij}\alpha_{ij}\lambda_i \lambda_j
+ \mathcal{O}(\bm{\lambda}^4).  \end{aligned} \label{eq:kscl}
\end{equation}

Comparing \req{eq:kscl} to Eqs.~(\ref{eq:k1}), (\ref{eq:kprime}), and (\ref{eq:k2}), the coefficients $\alpha_{ij}$ are nothing but the corresponding terms containing $\xi_{i}\xi_{j}$ in $K$.
Thus, we have

\begin{equation}
\alpha_{ij}=K_{ij}=\sum_{\sigma\sigma'} {K}_{ij}^{\sigma\sigma'}=\sum_{\sigma\sigma'} \widetilde{K}_{ij}^{\sigma\sigma'}.
\end{equation}
The SOC-scaling procedure is often used in DFT, probably due to its rather straightforward implementation.
Obviously, the scaling procedure can be generalized from sites to orbitals to obtain contributions from individual orbitals.

\subsection{Crystal structure of $L1_0$ compounds}

\begin{figure}[hbtp]
\includegraphics[width=0.7\linewidth,clip,angle=0]{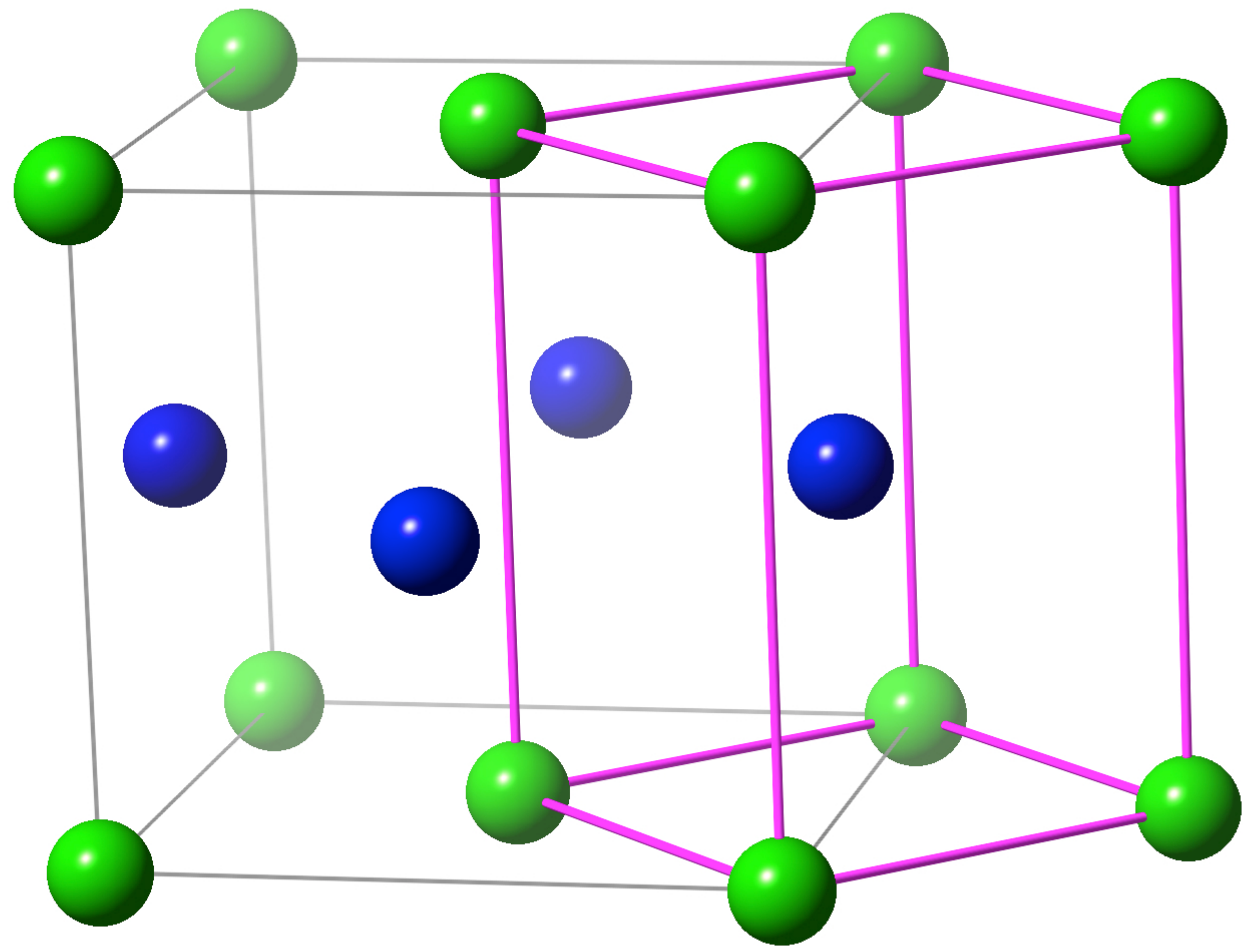}
\caption{Schematic representation of the CuAu-type $L1_0$ structure. }
\label{fig:xtal}
\end{figure}

We chose to focus on $L1_0$ magnetic compounds because they are one of the most widely studied systems~\cite{daalderop1991prb,solovyev1995prb,kota2014jpsj}.
Their simple CuAu-type crystal structure is shown in \rfig{fig:xtal}.
The primitive cell is body-centered tetragonal (\textit{bct}) and contains one formula unit (f.u.) while the conventional cell is face-centered tetragonal (\textit{fct}) and contains two f.u.
For all $L1_0$ magnetic compounds that we study in this work, experimental lattice parameters have been used.
The $c/a$ ratio values (with respect to the \textit{bct} primitive cell) are in the range of 1.28--1.414.
Considering that hypothetical \textit{bct}-FeCo structures with different $c/a$ ratios have received significant attention in the past few years, we also investigated a hypothetical \textit{bct}-FeCo structure with a $c/a$ ratio of 1.1.

\section{Results}
\label{sec:3}
\subsection{Electronic structure and magnetic moment}

\begin{widetext}
\begin{figure*}[hbt]
\begin{tabular}{c}
  \includegraphics[width=.99\linewidth,clip,angle=0]{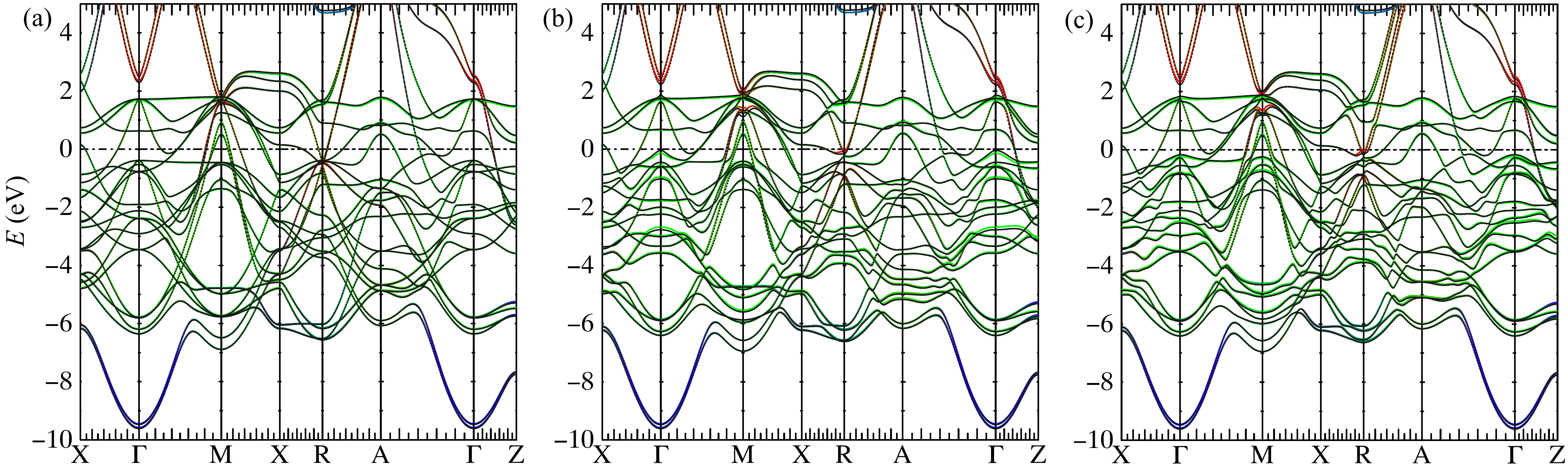}
\end{tabular}%
\caption{(a) Scalar-relativistic band structure of FePt using a tight-binding Hamiltonian and FP-LMTO (black dotted).
The tight-binding bands are in color, with blue identifying the $s$ states, red the (Fe-$p$, Pt-$p$) states, and green everything else.
The band structure with spin-orbit coupling is calculated with spin along the $[001]$ (b) and $[110]$ (c) directions.}
\label{fig:fept_band}
\end{figure*}
\end{widetext}

For all compounds we investigated in this work, the scalar-relativistic band structures recalculated in TB are essentially in perfect agreement with those obtained from DFT, within the specified energy window.
In both TB and DFT, the tetrahedron integration with Bl\"ochl correction method~\cite{blochl1994prb} is used to determine the Fermi level.
Here, we use FePt as an example to illustrate our implementation of SOC in the realistic TB framework.
The new Fermi level obtained in TB with charge neutrality deviates from the original DFT Fermi level by less than \SI{0.01}{\eV}.
As shown in \rfig{fig:fept_band}, without SOC, the band structures are nearly identical between TB bands and the all-electron full potential bands calculated using FP-LMTO.
The \textsc{vasp} bands (not shown) are essentially the same also.
The SOC constants $\xi_{il}$ calculated using FP-LMTO are used as input parameters to construct relativistic TB bands.

Directly using the SOC constants calculated in FP-LMTO, the resulting band structure's SOC splitting is overestimated just above and below the Fermi level at $R$ and $M$ points, respectively.
By investigating the eigenvectors of the non-SOC band in FP-LMTO and TB, we found that TB overestimates the eigenvector component of the Pt-$p$ states, which have a very large SOC constant of about \SI{2.6}{eV}.
(The site-and-orbital-resolved charges and moments are listed in Appendix~\ref{seca:qm}.)
This is because $p$ orbitals are much more extended than $d$ orbitals and we do not include the interstitial sites into the projection, and as a result, the interstitial components also fold into the atomic sites.
Thus, we simply renormalize the occupation numbers (or, equivalently, renormalize the SOC constants) of the Pt-$p$ channel using the ratio of atomic $p$ charge between \textsc{vasp} and TB.
This adjustment improves agreement between DFT bands and TB bands.
Correspondingly, the MAE also improves as we discuss later.
In comparison to a previously reported empirical TB method~\cite{zemen2014jmmm} that used Slater-Koster parametrization, our TB method shows significantly better agreement with DFT.

SOC parameters $\xi_{il}^{\uparrow\downarrow}$ calculated in FP-LMTO in various compounds are summarized in \rtbl{tbl:xi}.
Generally, $\xi$ integration is only significant near a nucleus where electrons move fast.
As a result, for a given element in various compounds, $\xi_l$ barely changes, especially for the $d$ channel, enabling transferability.
The $d$ orbitals of $3d$ elements are more spin polarized than those of $5d$ elements, and the ratio of $\xi_i^{\uparrow\uparrow}/\xi_i^{\uparrow\downarrow}\approx\xi_i^{\uparrow\downarrow}/\xi_i^{\downarrow\downarrow}$ of $3d$ elements varies between 1.07 and 1.13 in various $L1_0$ compounds.
For simplicity, we use the value of $\xi_{il}^{\uparrow\downarrow}$ for all spin channels.

\begin{table}[ht]
\caption{Spin-orbit coupling constants $\xi_i^{\uparrow\downarrow}$(meV) in various compounds calculated in FP-LMTO.
On the $3d$ sites (Mn, Fe, and Co),  $\xi_i^{\uparrow\uparrow}/\xi_i^{\uparrow\downarrow}\approx\xi_i^{\uparrow\downarrow}/\xi_i^{\downarrow\downarrow}$ varies between 1.07 and 1.13  for the $d$ channel.
SOC energies (isotropic) $\er$ (meV) calculated in TB are also listed.}
\label{tbl:xi}%
\bgroup
\def\arraystretch{1.1}%  1 is the default, change whatever you need
\begin{tabular*}{\linewidth}{l @{\extracolsep{\fill}} crrrrr}
\hline\hline
\multirow{2}{*}{Compound} & \multicolumn{2}{c}{1st element} & \multicolumn{2}{c}{2nd element} & \multirow{2}{*}{$\er$}\\ \cline{2-3} \cline{4-5}
         & $\xi_p$&$\xi_d$&$\xi_p$ & $\xi_d$    \\ \hline
    FePt &  197.2 &  55.0 & 2626.4 & 574.9  & 193.8  \\
    CoPt &  190.5 &  72.1 & 2793.1 & 580.8  & 210.6  \\
    FePd &  168.9 &  54.7 &  898.0 & 200.8  &  22.4  \\     
    FeNi &  232.9 &  55.9 &  218.2 &  91.6  &  11.7  \\     
    MnGa &  193.1 &  41.0 &  209.6 &  84.1  &   5.3  \\
    MnAl &  203.0 &  41.3 &   31.3 &   0.4  &   2.7  \\         
    \hline\hline
\end{tabular*}
\egroup
\end{table}

The SOC energy $\er$ is proportional to $\xi^2$ within second-order PT.
Among the $L1_0$ compounds we studied, CoPt has the largest $\xi$ values as well as the largest $\er$.
The $\er$ values are much larger than MAE, indicating that the isotropic part of $\er$ is much larger than the anisotropic part.
In comparison to other compounds, MnGa and MnAl have a rather high anisotropic/isotropic ratio, suggesting that they have a more ``efficient'' band structure, in the sense that the Fermi level is close to the bandfilling position that gives the largest MAE value, as we show in subsection~\ref{sec:result:bandfilling}. 
On the other hand, it is challenging to analyze the relationship between MAE and band structure for FePt and CoPt as each MAE value is only a small fraction of $\er$.

\subsection{MAE and band-filling effect}
\label{sec:result:bandfilling}

\begin{table}[ht]
\caption{ MAE ($\mu$eV/f.u.)
in $L1_0$ systems calculated in our tight-binding framework with (denoted as TB-PT) and without (denoted as TB) perturbation approach.
MAE values calculated using FP-LMTO with both PBE (denoted as FP) and BH (denoted as FP-LDA) exchange-correlation functionals, and using \textsc{vasp} with PBE functional, are also listed for comparison.
The SOC constants used in TB were obtained from FP using PBE functionals.}
\label{tbl:kall2}%
\bgroup
\def\arraystretch{1.1}%  1 is the default, change whatever you need
\begin{tabular*}{\linewidth}{l @{\extracolsep{\fill}} rrrrrr}
\hline
\hline
{Compound} & {TB}  & {FP}  & {TB-PT} & {VASP} & {FP-LDA} \\ \hline
 {FePt}    &  2495 & 2556  &  2692   & 2656   & 2746     \\ 
 {CoPt}    &   836 &  788  &  1098   &  858   & 1156     \\
 {FePd}    &   175 &  164  &   186   &  194   &  173     \\
 {FeNi}    &    66 &   68  &    65   &   80   &   68     \\
 {MnGa}    &   399 &  381  &   415   &  431   &  421     \\
 \hline\hline
\end{tabular*}
\egroup
\end{table}

As shown in \rtbl{tbl:kall2}, the MAE values calculated using TB agree well with DFT, especially with those obtained by the all-electron FP-LMTO method.
They are also comparable to previous DFT calculations using various methods~\cite{kota2014jpsj,antropov2014ssc}.
We found that MAE values calculated using PT generally agree very well with the MFT results.
The largest difference in MAE is in CoPt, in which PT gives a MAE 20\% larger than the MFT result.
Interestingly, the MAE of CoPt also strongly depends on the exchange-correlation functionals used.
LDA increases the MAE values by 30\%.
Likely, this is because CoPt has a large SOC and its MAE depends on the detailed, subtle band features near the Fermi level.
We will study CoPt in more detail in a later section.

\begin{figure}[htb]
\begin{tabular}{c}
\includegraphics[width=.98\linewidth,clip,angle=0]{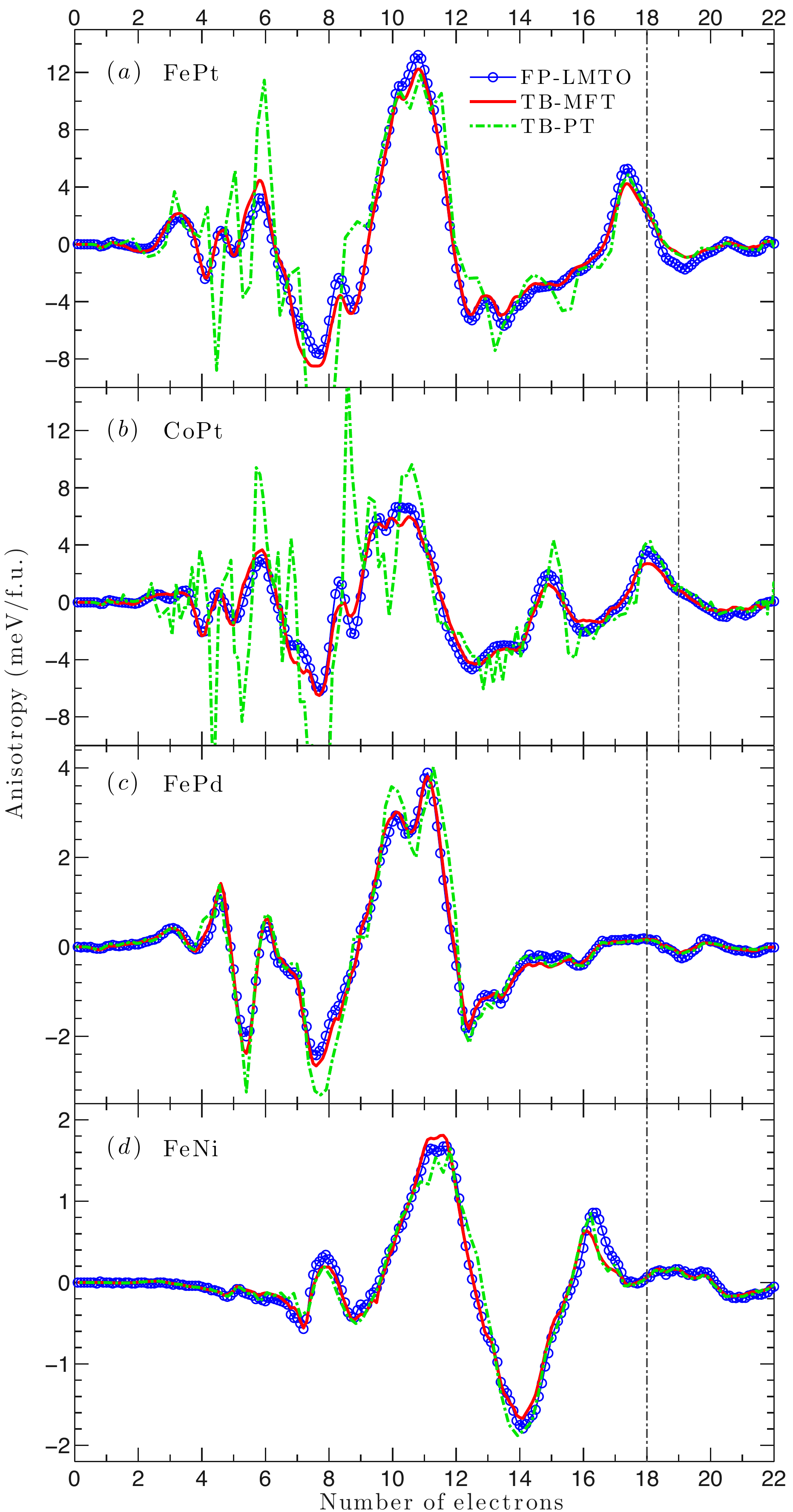}  \\
\end{tabular}%
\caption{MAE in FePt, CoPt, FePd, and FeNi as a function of band filling.
The solid and dashed lines are the results calculated using the magnetic force theorem and the second-order perturbation theory, respectively.
The blue dots are from FP-LMTO.
The vertical dashed-and-dotted lines indicate the actual number of valence electrons in each compound.}
\label{fig:k_vs_ne_fept_copt}
\end{figure}

\rFig{fig:k_vs_ne_fept_copt} shows the band-filling dependence of MAE in FePt, CoPt, FePd, and FeNi.
The oscillation~\cite{heine1984jmmm} of MAE can be associated with the local susceptibility~\cite{ke2015prb,inoue2015jap}.
For the entire bandfilling range, MAE values calculated using DFT and TB agree well.
On the other hand, although PT gives a good description of MAE at the actual electron filling for each compound, it disagrees with the MFT result in certain bandfilling ranges in FePt and CoPt.
Specifically, the disagreement is pronounced from four to eight electrons in FePt and from four to ten in CoPt.
A similar finding was reported by a previous LMTO-ASA study~\cite{kota2014jpsj}.
In comparing results for $3d$, $4d$, and $5d$ compounds, second-order PT is best suited for the lower SOC strength found in the lighter compounds, and that's where we obtained the best agreement between MFT and PT.
In general, we found that we could not improve agreement by using a denser $k$ mesh.

\begin{figure}[htb]
\begin{tabular}{c}
\includegraphics[width=.95\linewidth,clip,angle=0]{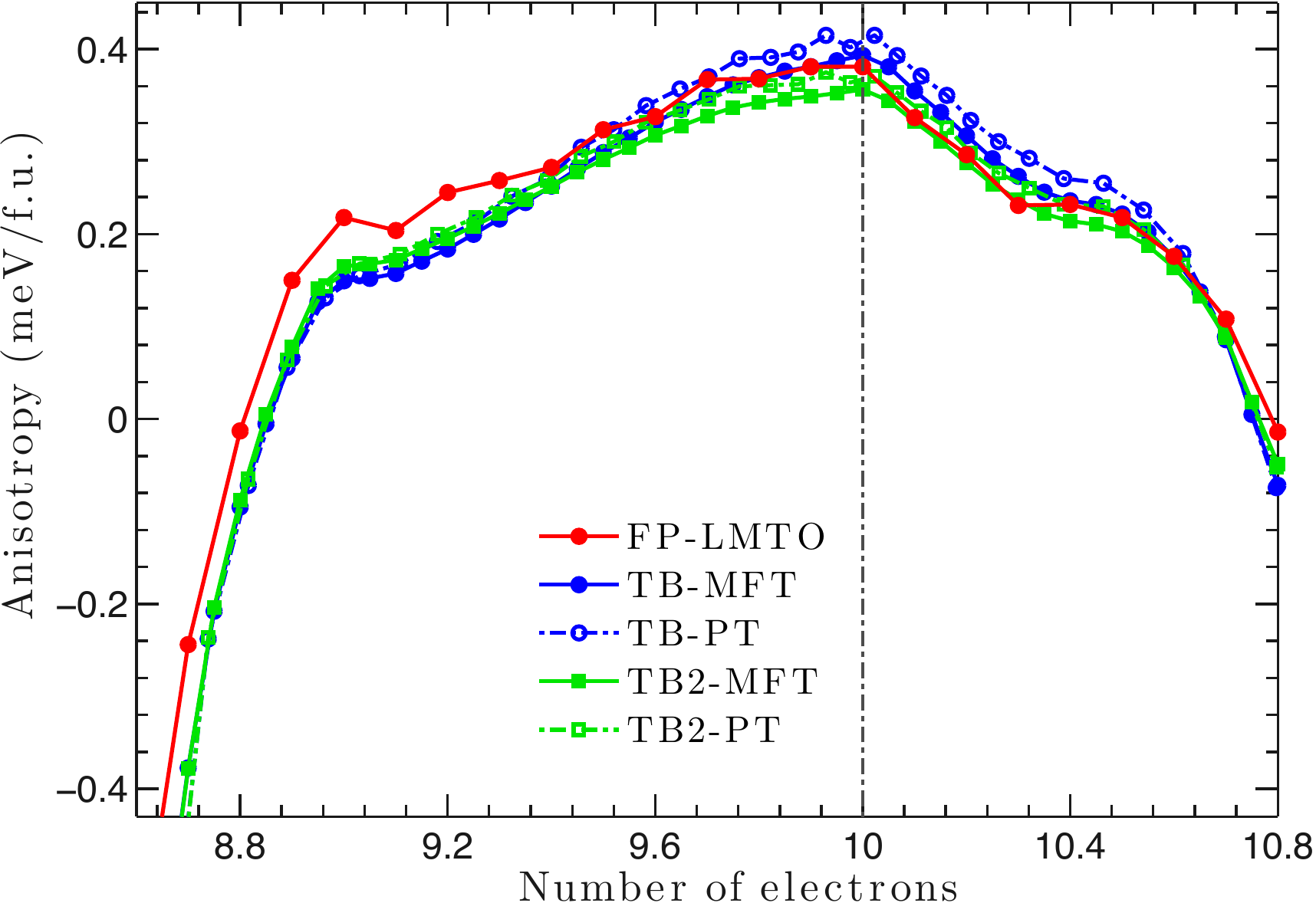} 
\end{tabular}%
\caption{ MAE in MnGa as a function of band filling calculated in TB and FP-LMTO.
The vertical dashed-and-dotted line at $N_\text{e}=10$ indicates the actual number of valence electrons in MnGa.
For TB calculations, we also consider the renormalization of Ga-$p$ and Mn-$p$ electron occupancy when including SOC (denoted as TB2).}
\label{fig:k_vs_ne_mnga_mnal}
\end{figure}

Although MnGa does not contain heavier $4d$ and $5d$ elements, its MAE is larger than FePd.
As shown in \rfig{fig:k_vs_ne_mnga_mnal}, the Fermi level is located close to the filling with maximum MAE.
PT generally agrees well with MFT.
We also consider the renormalization of the $p$ (for both Ga and Mn) charge when including SOC, and it slightly decreases the MAE.
All TB results are within $\pm10\%$ of the FP-LMTO results.
While MnGa MAE is about \SI{10}{\percent} of the SOC energy, the ratio is much higher than in FePt and CoPt.

\subsection{Reciprocal-space resolved MAE and its correlation with Fermi surface}

\begin{figure}[thb]
  \includegraphics[width=.95\linewidth,clip,angle=0]{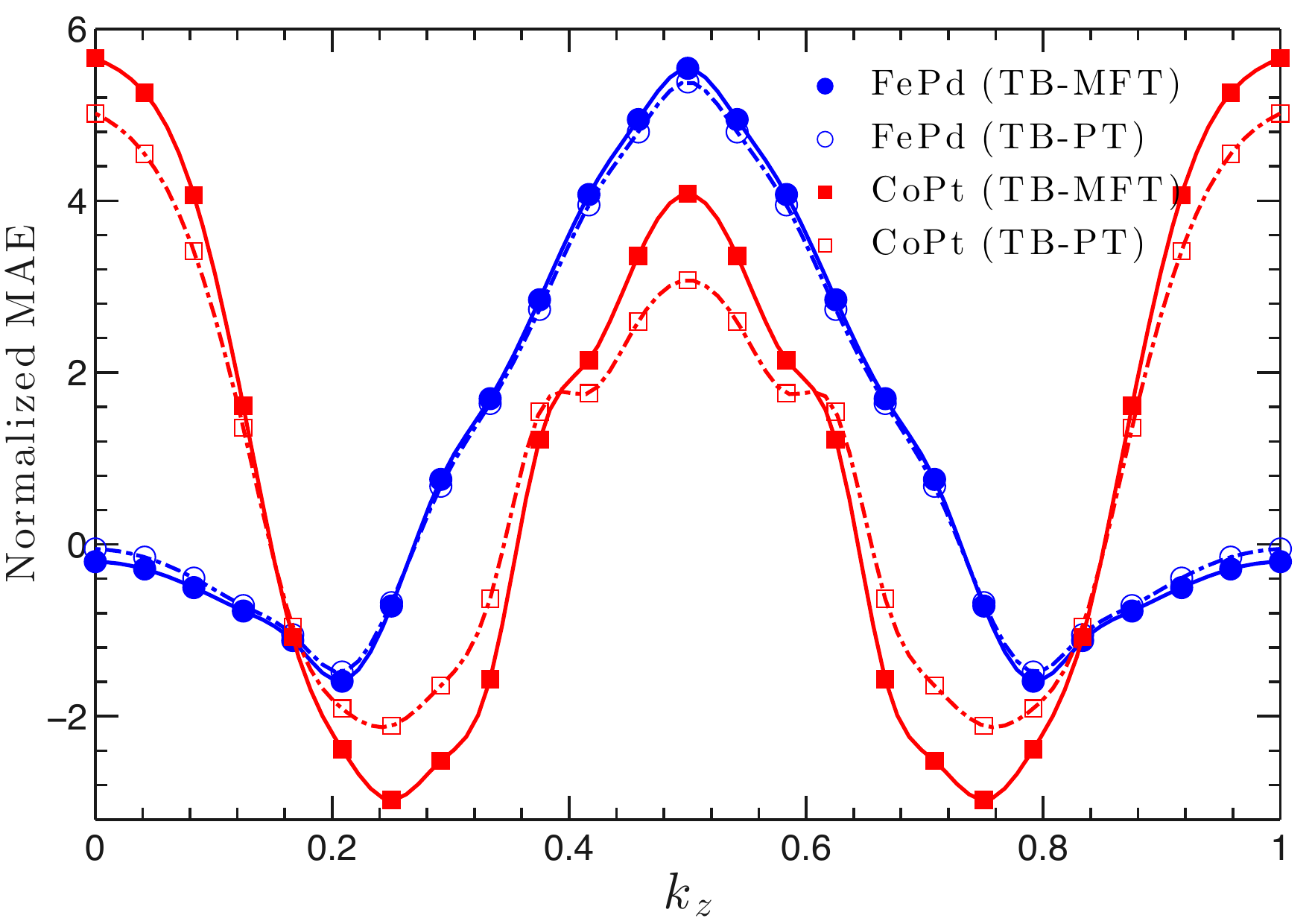}
\caption{$k_z$-resolved normalized MAE of FePd and CoPt calculated in TB.
Calculations are performed with and without using perturbation theory.
For the sake of comparison, MAE is normalized so that the average contribution from each $k_z$ plane is equal to 1.}
\label{fig:mae_vs_kz_fepd_copt}
\end{figure}

As shown in \rtbl{tbl:kall2}, MAE values calculated using second-order TB with either MFT or PT generally agree well with each other and with the corresponding DFT calculations.
To further validate the applicability of PT, we compare the $k$-resolved MAE using both methods for a more stringent test.
Remarkably, for all $L1_0$ compounds we study, the two methods produce very similar results, further suggesting that the overall effect of Fermi surface relaxation is non-significant and that second-order PT is valid.
CoPt has the largest difference of MAE values between MFT and PT among the compounds we study.
\rFig{fig:mae_vs_kz_fepd_copt} shows the $k_z$ dependence of MAE contributions calculated in FePd and CoPt.
Although the $k_z$-resolved MAE in CoPt shows a larger difference, the two methods still generally agree with each other.
The larger difference is likely due to its rather large SOC for the Pt atom and more complex Fermi surface, which results in a larger Fermi surface relaxation effect.
Similar to FePd, the $k_z$-resolved MAE calculated using MFT and PT in other $L1_0$ compounds (not shown) are also nearly identical.
And, the $Z$-$R$-$A$ ($k_z=0.5$) plane has the largest positive contribution to MAE.

\begin{figure}[bht]
  \includegraphics[width=.95\linewidth,clip,angle=0]{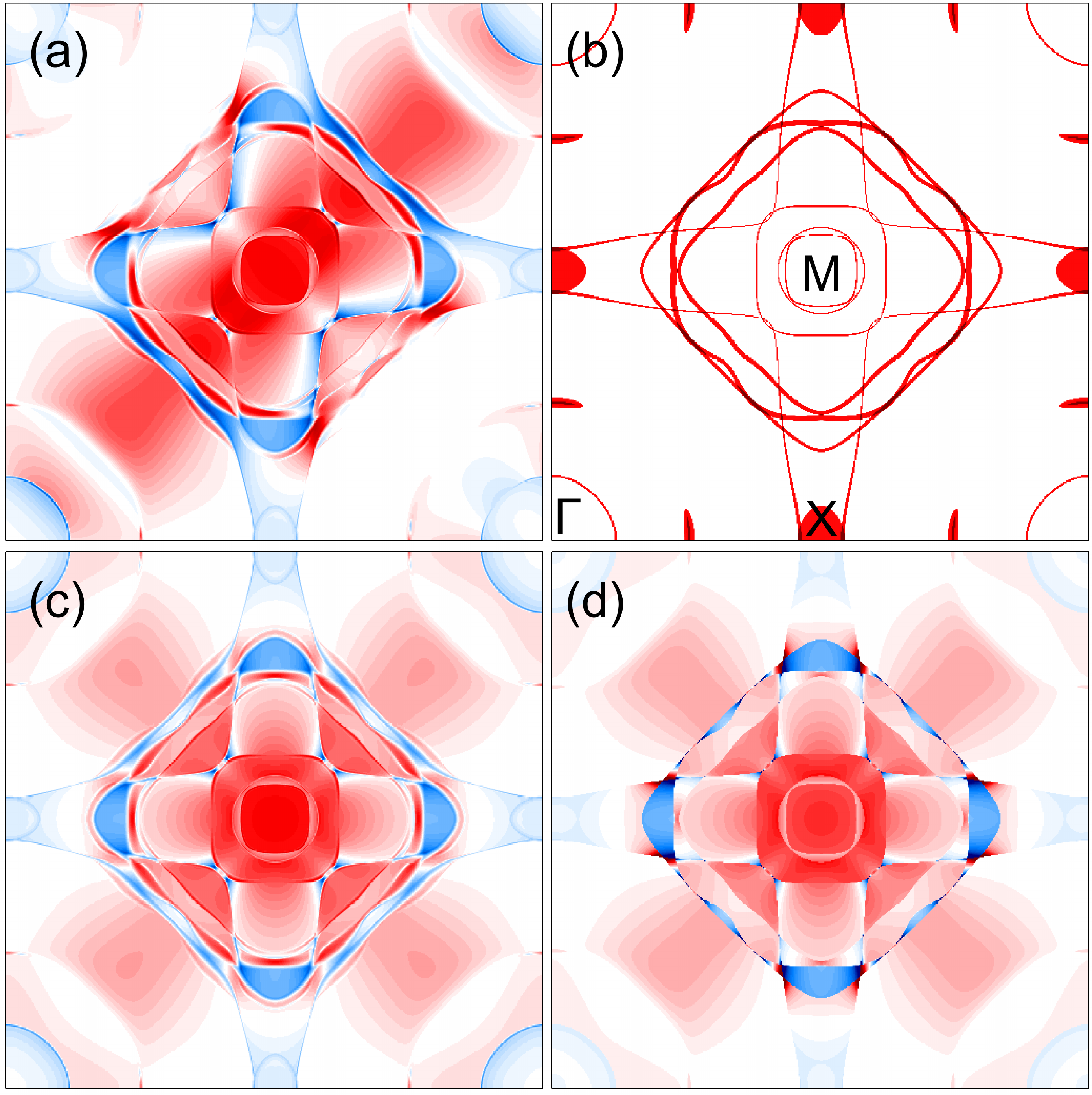}
\caption{$k$-resolved MAE and Fermi surface contour in CoPt for $k_z=0$ calculated in TB.
Red (blue) color indicates positive (negative) contributions to MAE.
(a) $k$-resolved MAE calculated via the magnetic force theorem.
(b) Fermi surface contour plot.
(c) Symmetrized $k$-resolved MAE.
(d) Symmetrized $k$-resolved MAE calculated via perturbation theory.
The number of $k$ points used in the full Brillouin zone is $\sim10^8$.}
\label{fig:mae_kxky_resolved_copt}
\end{figure}
The $(k_{x},k_y)$-resolved MAE at $k_z=0$ in CoPt calculated using both MFT and PT are shown in \rfig{fig:mae_kxky_resolved_copt}, which also includes the corresponding non-SOC Fermi contour.
\rFig{fig:mae_kxky_resolved_copt}(a) shows the resolved MAE calculated using MFT.
Obviously, the four-fold symmetry is broken due to SOC.
The twofold or mirror symmetry along the [110] direction is simply an artifact of our choice of [110] as the reference in-plane spin direction.
The Fermi contour plot, as shown in \rfig{fig:mae_kxky_resolved_copt}(b), is calculated by integrating the electron density at the Fermi level with a width of \SI{0.02}{\eV}.
To better compare with the Fermi surface and $k$-resolved MAE, we also symmetrized $k$-space contributions with symmetry operations that are compatible with SOC when the spin quantization axis is along the $z$ direction.
The symmetrized $k$-resolved MAE calculated in MFT and PT are shown in Figs.~\ref{fig:mae_kxky_resolved_copt}(c) and \ref{fig:mae_kxky_resolved_copt}(d), respectively.

As shown in Figs.~\ref{fig:mae_kxky_resolved_copt}(b) and \ref{fig:mae_kxky_resolved_copt}(d), the correlation between the $(k_{x},k_y)$-resolved MAE and the Fermi contour is apparent and two features stand out.
First, large changes of MAE contributions occur at the Fermi contour.
This is because MAE contributions are obtained by integrating \req{eq:k2} up to $\epsilon_\text{F}$.
When the $k$ path crosses the Fermi contour, some bands become occupied or unoccupied, and their MAE contributions appear or disappear.
Second, from the point of view of PT, the strongest contributions to MAE are from those virtual transitions between the unoccupied and occupied states near the Fermi level that are coupled by SOC~\cite{ke2015prb,daalderop1994prb}.
As a result, the largest contributions are located at and near the degenerate states across the Fermi contour, as expected from PT, and as shown in \rfig{fig:mae_kxky_resolved_copt}(d).

As shown in Figs.~\ref{fig:mae_kxky_resolved_copt}(c) and \ref{fig:mae_kxky_resolved_copt}(d), the overall $(k_{x},k_y)$-resolved MAE calculated using the two methods share great similarity.
The largest differences exist around the Fermi contour, where the relaxation effect is large.
The large contributions (near the degenerate states) observed in PT disappear in MFT.
Indeed, the MFT results show more complex features than the non-SOC Fermi contour, which corresponds to the Fermi surfaces when SOC is on and the magnetization direction being along in- and out-of-plane directions. The SOC-induced lifting of band degeneracy, especially near the $\epsilon_\text{F}$, is often discussed to explain MAE in various systems~\cite{wang1993prb,daalderop1994prb,moos1996ssc,ravindran2001prb}. 

Thus, we demonstrate that MFT and PT give very similar results for not only total MAE but also for $k$-resolved MAE.
To achieve this, it is important to use the reference energy $\epsilon^{0}_\text{F}$ as in \req{eqn_k_esum_ef0}.
Otherwise, if one just resolves MAE using \req{eqn_k_esum}, a very different dependence can be obtained, and $k$-resolved MAE will only manifest the Fermi surface, near which the change of band occupancy is significant.
MAE has often been resolved into $k$ space along certain line paths between high-symmetry points.
Not surprisingly, using \req{eqn_k_esum} to resolve MAE will result in spikes at points where bands cross the Fermi level.
Further resolution of MAE into atomic sites by projecting eigenvectors of each $k$ point may produce unphysical results~\cite{subkow2009prb,aberg2015tech-osma,sipr2014jpcm}.

\subsection{Two-ion MA: Intersublattice contribution}

Site-resolved MAE values, together with exchange parameters, can be used to construct an atomic spin Hamiltonian for subsequent Monte Carlo or spin-dynamics simulations to calculate the temperature dependence of magnetic properties.
Methods such as evaluating the anisotropy of on-site SOC energy~\cite{antropov2014ssc}, which is a local quantity, have been used to resolve the MAE contribution from each individual sublattice.
Here, we use PT in TB to resolve MAE into sublattices and validate the decomposition using the SOC-strength-scaling approach in \textsc{vasp}.

\begin{table}[htb]
\caption{Sublattice-resolved MAE (normalized to 1) in $L1_0$ systems calculated using perturbation theory in TB.
For each compound $AB$, MAE is resolved into intrasublattice contributions, $K_{A\text{-}A}$ and $K_{B\text{-}B}$, and intersublattice contribution $K_{A\text{-}B}$.
We also define the contribution from individual sublattice $A$ as $K_A=(K_{A\text{-}A}+K_{A\text{-}B}/2)$.
The hypothetical FeCo structure with $c/a=1.1$ is also included.}
\label{tbl:k_sublattice_tbpt}%
\bgroup
\def\arraystretch{1.1}%  1 is the default, change whatever you need
\begin{tabular*}{\linewidth}{l @{\extracolsep{\fill}} rrrrr}
\hline
\hline
$AB$ & $K_{A\text{-}A}$  & $K_{A\text{-}B}$ & $K_{B\text{-}B}$ & $K_A$ & $K_B$ \\ 
\hline
FePt & 0.11 & -0.55 &  1.44 & -0.16 &  1.16  \\
CoPt & 0.17 & -0.77 &  1.59 & -0.21 &  1.21  \\
FePd & 1.80 & -2.58 &  1.78 &  0.51 &  0.49  \\
FeNi & 4.51 & -5.88 &  2.37 &  1.57 & -0.57  \\
MnGa & 0.66 &  0.24 &  0.10 &  0.78 &  0.22  \\
MnAl & 0.98 &  0.02 &  0.00 &  0.99 &  0.01  \\
FeCo & 0.10 &  0.63 &  0.28 &  0.41 &  0.59  \\
\hline
\hline
\end{tabular*}
\egroup
\end{table}

As discussed above, PT can well describe the MAE in these systems.
To quantify the single-ion and two-ion contributions of MAE, we first use PT within TB to resolve MAE into intrasublattices and intersublattice contribution.
Results are summarized in \rtbl{tbl:k_sublattice_tbpt}.
Interestingly, all intrasublattice contributions are positive in all elements except for the $s$-like Al site in MnAl, where it vanishes.
The sign of the intersublattice contribution varies and its amplitude is generally comparable to or even larger than that of either individual intrasublattice.
For FePt and CoPt, the major contributions are from the Pt sites.
The intersublattice contributions are negative for FePt, CoPt, FePd, and FeNi.
Especially in FeNi, the amplitude of the negative intersublattice contribution is larger than each individual intrasublattice contribution.
In contrast, the intersublattice contribution is positive in MnGa.
An even larger positive intersublattice contribution is found in hypothetical FeCo with $c/a=1.1$.

\begin{figure}[thb]
\begin{tabular}{cc}
\includegraphics[width=.47\linewidth,clip,angle=0]{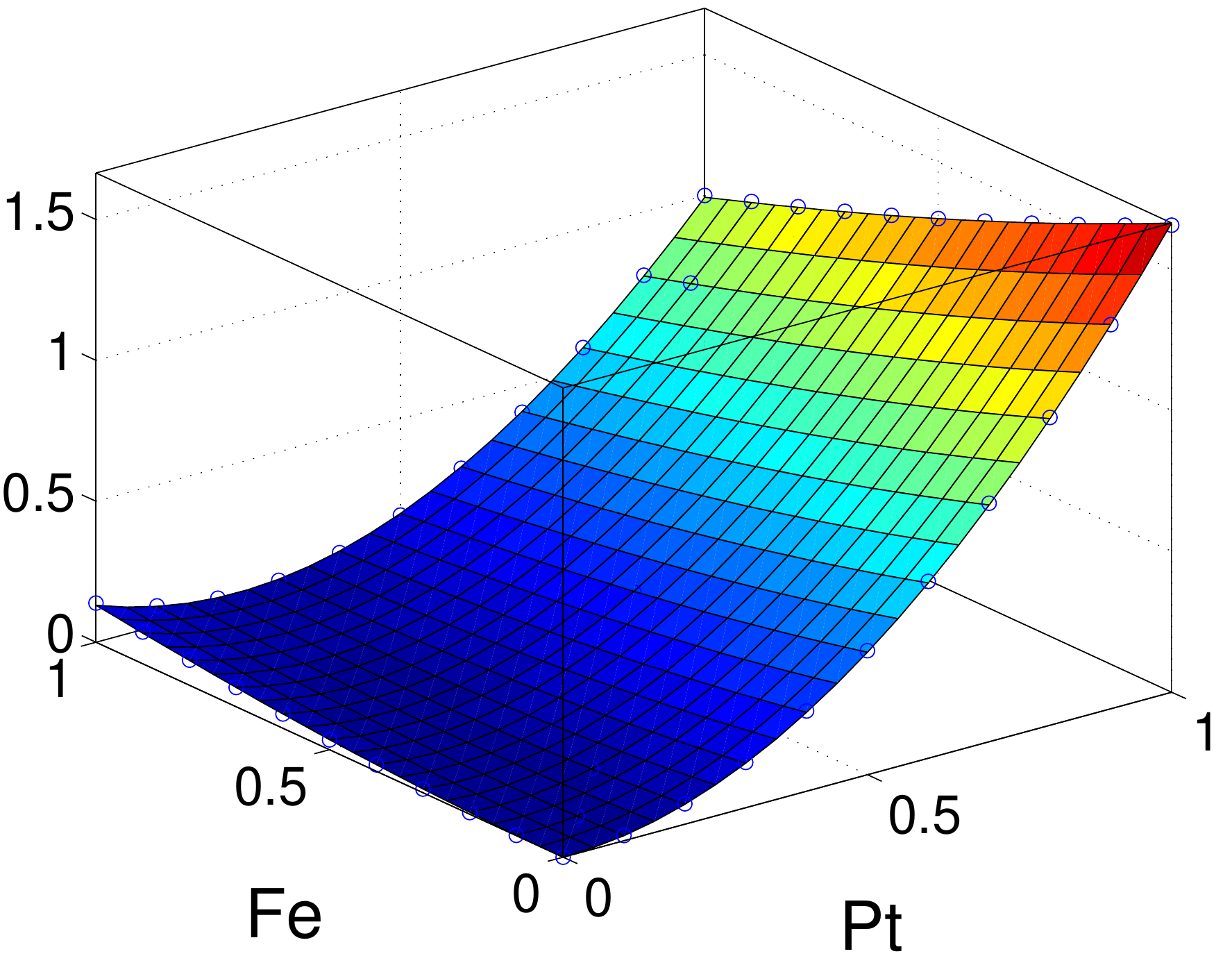} & %
\includegraphics[width=.47\linewidth,clip,angle=0]{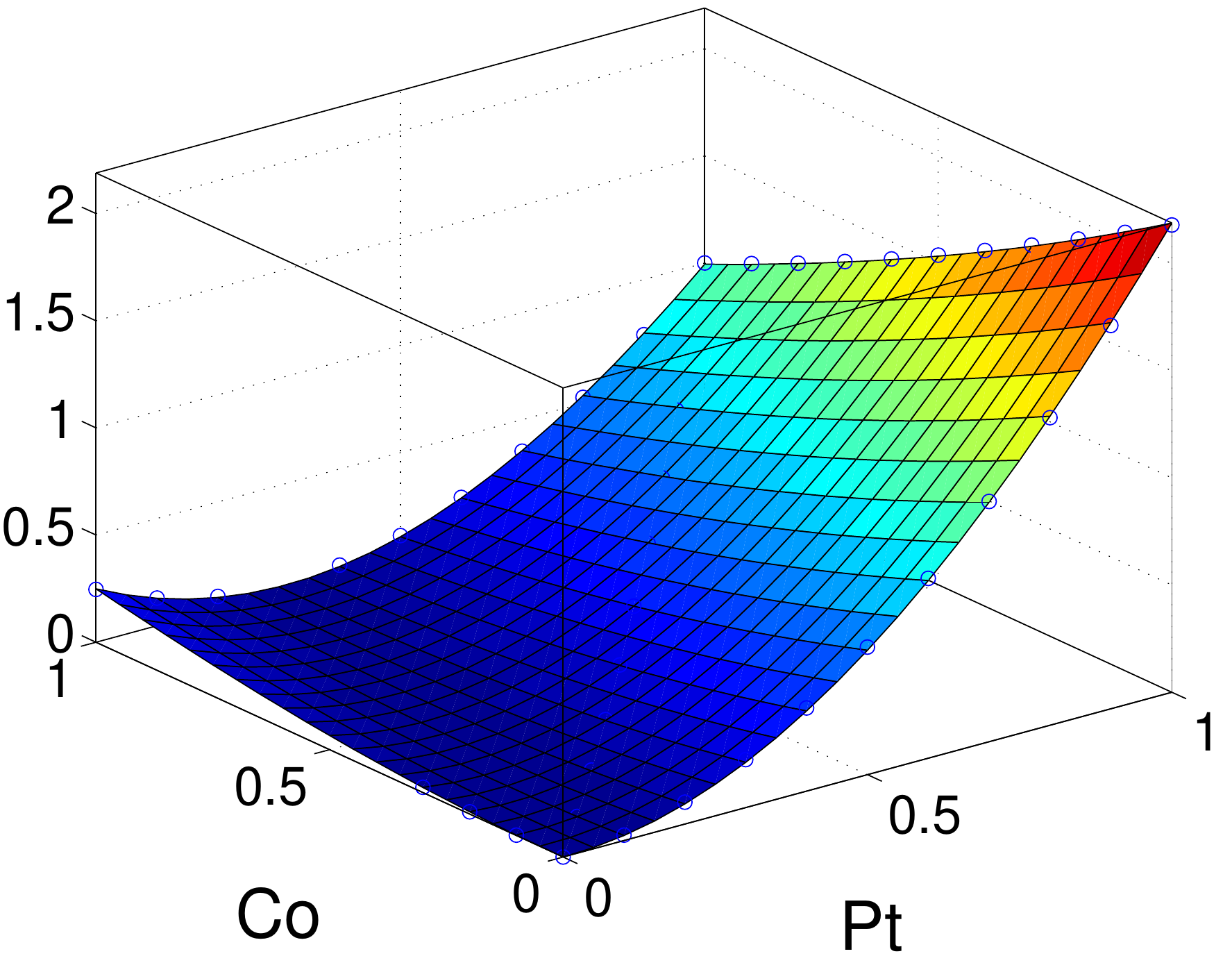} \\
\includegraphics[width=.47\linewidth,clip,angle=0]{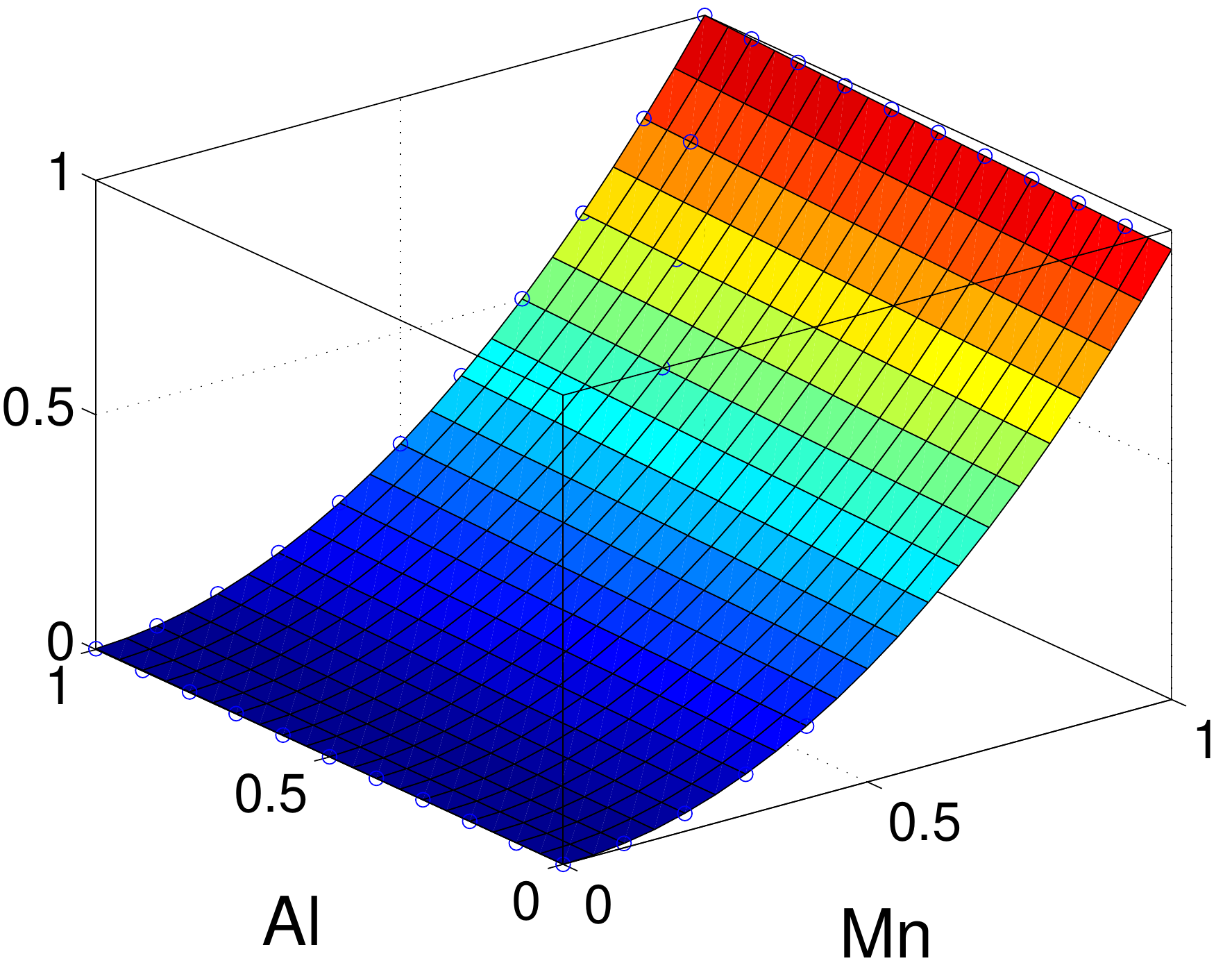} & %
\includegraphics[width=.47\linewidth,clip,angle=0]{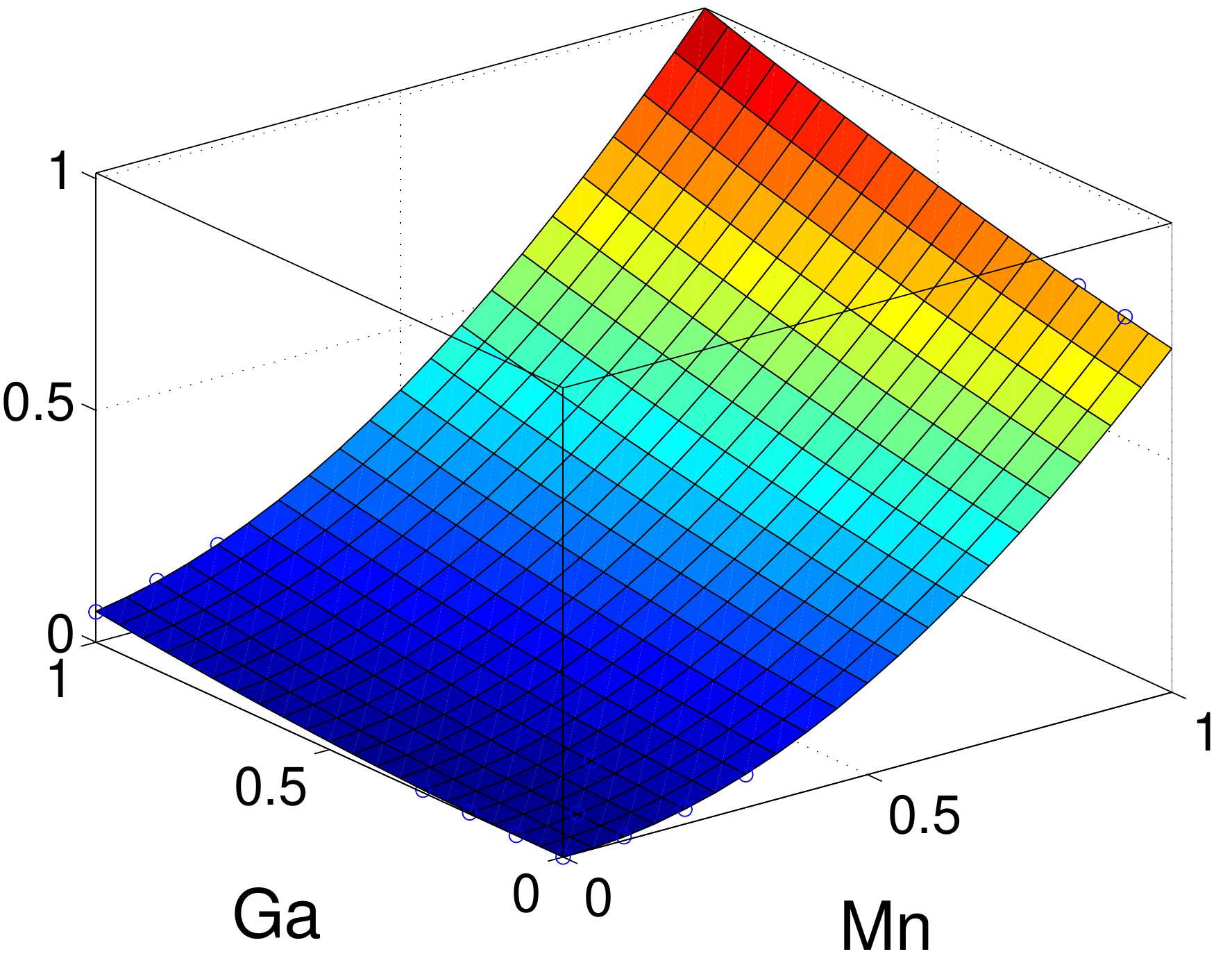} \\
\includegraphics[width=.47\linewidth,clip,angle=0]{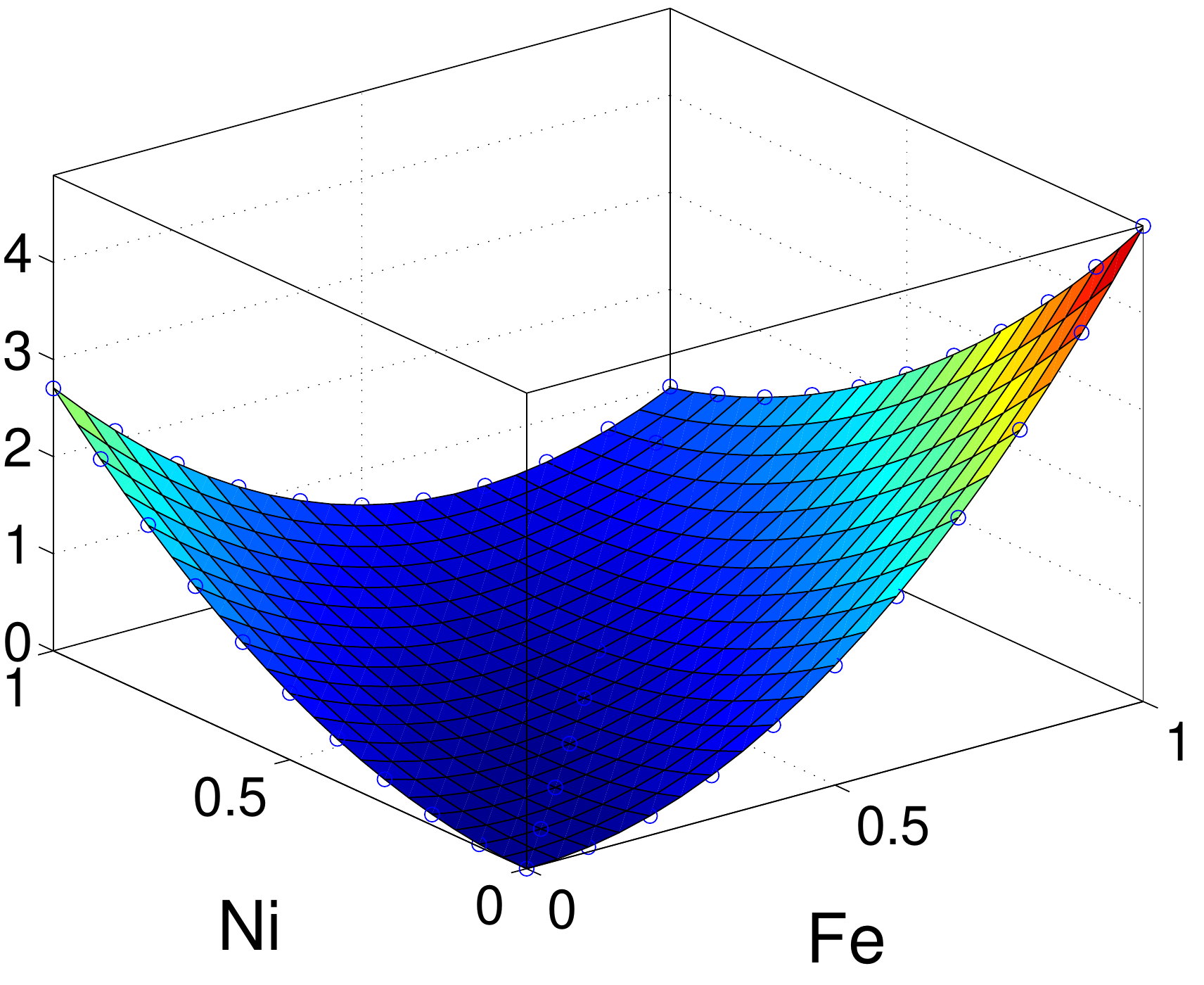} & %
\includegraphics[width=.47\linewidth,clip,angle=0]{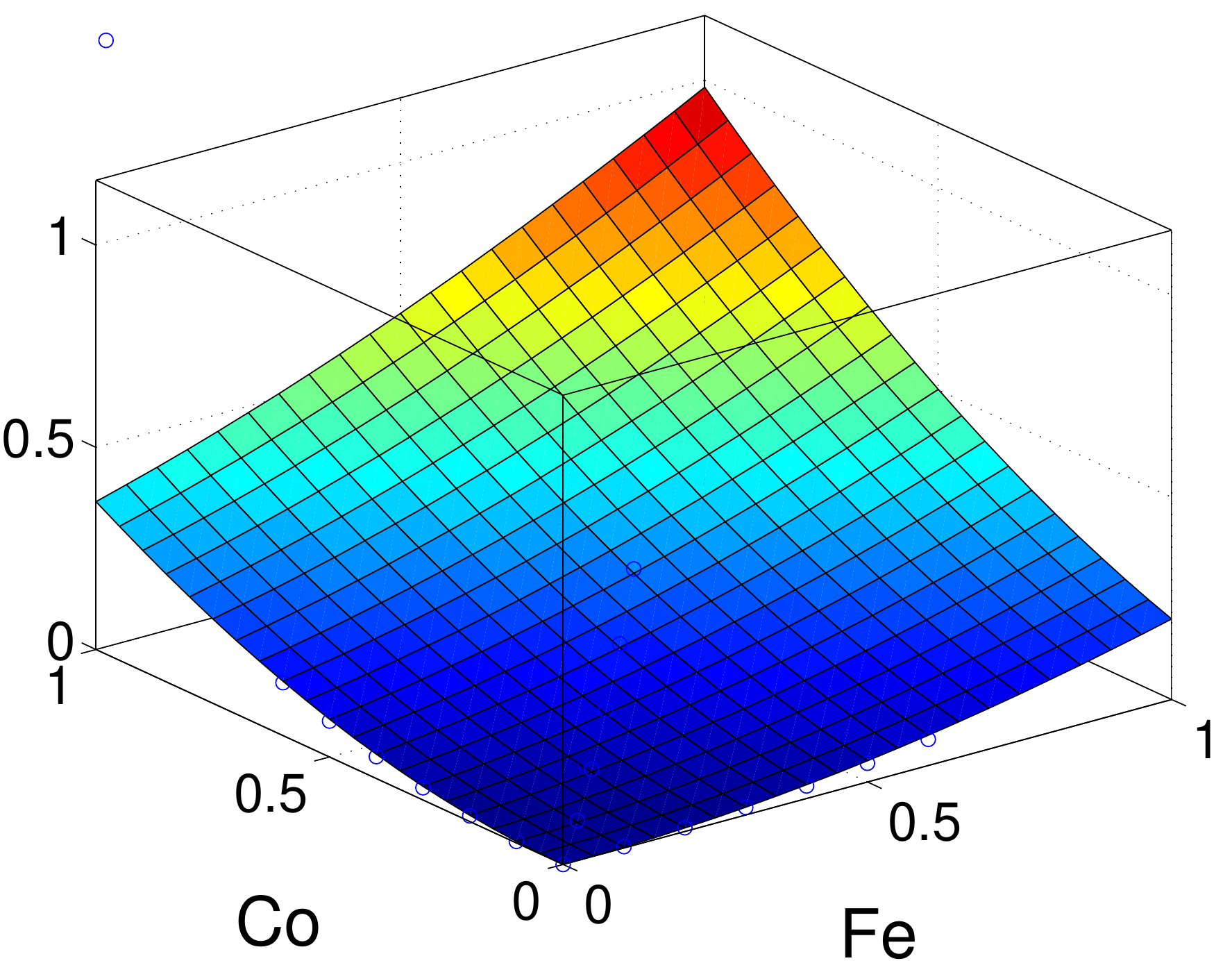}
\end{tabular}%
\caption{Normalized MAE $K(\lambda_i,\lambda_j)/K(\lambda_i=1,\lambda_j=1)$ in $L1_0$ compounds as a function of SOC-scaling factors $\lambda_i$ and $\lambda_j$ calculated using second-variation method in \textsc{vasp}.
The SOC strengths are scaled between 0 and 1.
}
\label{fig:ktot_soc}
\end{figure}

To validate TB results, we also investigate the intrasublattice and intersublattice MAE contributions by scaling the SOC strength in \textsc{vasp}.
\rFig{fig:ktot_soc} shows the normalized MAE as a function of the SOC-scaling factors (between 0 and 1) for $L1_0$ materials using the second-variation method in \textsc{vasp}.
For all compounds, the sign and relative amplitude of intrasublattice and intersublattice contributions agrees well with TB-PT results.
Furthermore, we fit MAE as a function of SOC-scaling factors (with $0.9<\lambda_i<1.1$) using \req{eq:kscl} and further confirm that the second-order terms agree very well with TB results listed in \rtbl{tbl:k_sublattice_tbpt}.
The fourth-order terms are generally small especially for $4d$ and $3d$ compounds.
Owing to stronger SOC, FePt and CoPt have larger and negative fourth-order contributions: $\sim$8\% of total MAE.
A previous study also found a small and negative high-order contribution to MAE in FePt~\cite{ayaz-khan2016prb}.
The good agreement between TB and \textsc{vasp} further validates the accuracy of the PT approach for those systems.
Unlike the scaling procedure, the PT approach resolves all contributions in a single calculation.
Thus, for the same analysis, once the TB Hamiltonian is constructed, the TB-PT approach is orders of magnitude faster than the SOC-scaling approach in DFT.

We further investigate FeNi, in which the intersublattice MAE dominates.
Along each direction, the intrasublattice SOC energy is large and dominated by the $\lambda^2$ term while the intersublattice term is rather small.
However, the majority of intrasublattice terms cancel out between the two directions, while the intersublattice term does not.
Hence, the intersublattice term becomes dominant in the anisotropy of SOC.
In other words, the intrasublattice terms in SOC are large but more isotropic, while the intersublattice terms are smaller but more anisotropic with respect to the magnetization quantization direction.

A large intersublattice MAE contribution may suggest the need to go beyond the single-ion MAE model when interfacing $\abinitio$ methods with atomic spin simulation.
To simulate temperature-dependent MAE or other magnetic properties, exchange coupling is often included over the first one or two nearest-neighbor shells while MAE is often included using the single-ion MAE term $k_{i}(s^z_i)^2$.
For systems with strong intersublattice contribution, one may also need to include two-ion terms such as $k_{ij}s^z_is^z_j$ into the atomic spin Hamiltonian.

\section{Conclusions}
Using $L1_0$ systems as a test case, we demonstrate that the {\abinitio} TB framework, constructed using the maximally localized Wannier functions method, can be used to efficiently and accurately compute and resolve MAE in transition metal systems.
With the magnetic force theorem, TB quantitatively reproduces DFT results over the full band-filling range from the bottom of valence band to a few eV above the Fermi level.
When calculating $k$-resolved MAE in TB, the magnetic force theorem and perturbation theory results agree with one another, and both yield MAE contour maps that are consistent with the Fermi surface.
We also resolve MAE into intrasublattice and intersublattice contributions using perturbation theory in TB and a scaled spin-orbit strength procedure in DFT.
The results using these two methods are in excellent agreement.
We found that the sign of the intersublattice contribution differs among compounds, and its amplitude may be comparable to or even larger than the intrasublattice contributions, suggesting the need to go beyond the single-ion MAE model.
Depending on the system size, once the TB Hamiltonian is constructed, it can speed up the calculation by orders of magnitude, providing an efficient, accurate, and high-resolution method to calculate MAE.
We expect that it can be applied to more complex compounds and structures to compute and analyze MAE.
Finally, this realistic TB method can also be interfaced with {\abinitio} methods beyond DFT, such as the much more expensive self-consistent $GW$ methods~\cite{van-schilfgaarde2006prl,kotani2007prb}, to greatly accelerate the calculations and analysis of MAE or other SOC-related properties using those methods.

\section*{Acknowledgments}
I thank B.~Harmon for helpful discussions.
This work was supported by the U.S.~Department of Energy, Office of Science, Office of Basic Energy Sciences, Materials Sciences and Engineering Division, and Early Career Research Program.
The initial development of the software was supported by the Laboratory Directed Research and Development Program of Ames Laboratory.
Ames Laboratory is operated for the U.S.~Department of Energy by Iowa State University under Contract No.~DE-AC02-07CH11358.

\appendix
\section{Spin-orbit coupling operators in real spherical harmonics representation}
\label{seca:soc}
\begin{equation}
\mathbf{L} \cdot \mathbf{S}
={\frac{1}{2}}(\mathbf{J}^{2}-\mathbf{L}^{2}-\mathbf {S} ^{2})={\frac
  {\hbar
    ^{2}}{2}}{\begin{pmatrix}L_{z}&L_{-}\\L_{+}&-L_{z}\end{pmatrix}}.
\end{equation}

In the presentation of complex spherical harmonics $Y_{\ell }^{m}$, the non-vanished matrix elements of $L_{z}$, $L_{+}$, and $L_{-}$ are
\begin{eqnarray}
  \langle l,m   | L_z |l,m \rangle &=& \hbar m \nonumber, \\ 
  \langle l,m-1 | L_- |l,m \rangle &=& \hbar \sqrt{l(l+1)-m(m-1)},  \\
  \langle l,m+1 | L_+ |l,m \rangle &=& \hbar \sqrt{l(l+1)-m(m+1)}.  \nonumber
\end{eqnarray}  

\textsc{wannier90} uses the real spherical harmonics $Y_{\ell m}$, also known as tesseral spherical harmonics, which can be written in terms of the complex spherical $Y_{\ell }^{m}$ as
\begin{equation}
\label{eq:math-mapping}
{\begin{aligned}
Y_{\ell m}&=
{\begin{cases}\displaystyle
    {i \over {\sqrt {2}}}\left(Y_{\ell }^{-|m|}-(-1)^{m}\,Y_{\ell }^{|m|}\right)&{\text{if}}\ m<0,\\
    \displaystyle Y_{\ell }^{0}&{\text{if}}\ m=0,\\
    \displaystyle {1 \over {\sqrt {2}}}\left(Y_{\ell }^{-|m|}+(-1)^{m}\,Y_{\ell }^{|m|}\right)&{\text{if}}\ m>0.
\end{cases}}
\end{aligned}}
\end{equation}

The angular momentum matrices in the real-spherical-harmonics representation, ${\bf O}^{(\mathbb{R})}$, can be obtained by directly evaluating the angular momentum operator on real spherical functions in \req{eq:math-mapping}, or transforming ${\bf O}^{(\mathbb{C})}$, the corresponding operator matrix from complex representation:

\begin{eqnarray}
{\bf O}^{(\mathbb{R})} &=& {\bf U}_{\mathbb{R}\leftarrow\mathbb{C}} {\bf O}^{(\mathbb{C})} {\bf U}_{\mathbb{R}\leftarrow\mathbb{C}} ^\dagger
={\bf U}_{\mathbb{R}\leftarrow\mathbb{C}} {\bf O}^{(\mathbb{C})} {\bf U}_{\mathbb{C}\leftarrow\mathbb{R}}.
\end{eqnarray}

From \req{eq:math-mapping}, the transfer matrix ${\bf U}_{\mathbb{C}\leftarrow\mathbb{R}}$ can be written as

\begin{equation}
{\bf U}_{\mathbb{C}\leftarrow\mathbb{R}} =
\frac{1}{\sqrt{2}}\left(
\begin{array}
[c]{ccccccc}%
i &  0 &  0 & 0 &  0 & 0 & 1\\ 
0 &  i &  0 & 0 &  0 & 1 & 0\\
0 &  0 &  i & 0 &  1 & 0 & 0\\
0 &  0 &  0 & \sqrt{2} & 0 & 0 & 0\\
0 &  0 &  i & 0 & -1 & 0 & 0\\
0 & -i &  0 & 0 &  0 & 1 & 0 \\
i &  0 &  0 & 0 &  0 & 0 &-1\\ 
\end{array}
\right).
\label{eq:ucr}
\end{equation}
For $p$ and $d$ orbitals, only the corresponding subblocks of \req{eq:ucr} are needed.

Similarly, for quantization along other directions, the matrix can be rotated by using Wigner matrix
\begin{equation}
H_\text{so}(\hat{\bf n})=\frac{\xi}{2}U(\theta,\varphi)(\mathbf{L}
\cdot \mathbf{S}) U^\dagger(\theta,\varphi),
\end{equation}

and

\begin{equation}
U\left(\theta,\varphi\right)=
\left(
\begin{array}{cc}
e^{i\frac{\phi }{2}} \cos \left(\frac{\theta }{2}\right) & e^{-i\frac{\phi }{2}} \sin \left(\frac{\theta
}{2}\right) \\
-e^{i\frac{\phi }{2}} \sin \left(\frac{\theta }{2}\right) & e^{-i\frac{\phi }{2}} \cos \left(\frac{\theta
}{2}\right) \\
\end{array}
\right),
\end{equation}

where $\theta$ and $\varphi$ are the angles of the direction of magnetization when the unit vector is defined by $\hat{\bf n}=(\sin\theta\cos\varphi, \sin\theta\sin\varphi, \cos\theta)$.
When the spin quantization axis is along [110] direction, the SOC Hamiltonian can be written as

\begin{equation}
\begin{gathered}
  H_\text{so}(\uvecn_{110}) \\
  =\frac{\xi}{2}
\left(
\begin{array}{cc}
\frac{\sqrt{2}}{2} \left({L_x+L_y}\right) &
-L_z + \frac{i\sqrt{2}}{2} \left(L_x-L_y\right)\\
-L_z - \frac{i\sqrt{2}}{2} \left(L_x-L_y\right) &
-\frac{\sqrt{2}}{2} \left({L_x+L_y}\right) \\
\end{array}
\right).
\end{gathered}
\end{equation}

\section{Charge and moment calculated in TB and VASP}
\label{seca:qm}
\rtbl{tbl:mom} lists the site-resolved charge and magnetic moments in $L1_0$ compounds calculated in TB and \textsc{vasp}.
\begin{widetext}
\begin{table*}[htb]
\caption{Site-and-orbital-resolved charge $q$ and magnetic moment $m$ ($\mu_\text{B}$/atom) in $L1_0$ compounds calculated in TB and \textsc{vasp}.
The moments of $s$-orbitals are negligible and not shown. Spin-orbit is not included in calculation. }
\label{tbl:mom}%
\centering
\bgroup
\def\arraystretch{1.1}%  1 is the default, change whatever you need
\begin{tabular*}{\linewidth}{lc@{\extracolsep{\fill}}rrrrrrrrrrrrrrrrrrrrc}
\hline\hline
%% \multirow{2}{*}{Compound} & \multirow{2}{*}{Method} & \multicolumn{7}{c}{$1_{st}$ element} & \multicolumn{7}{c}{$2_{nd}$ element} & \multicolumn{2}{c}{\mw{Cell}}\\ \cline{3-9} \cline{10-16} 
%% &        & \multicolumn{1}{c}{$s$} & \multicolumn{2}{c}{$p$} & \multicolumn{2}{c}{$d$} & \multicolumn{2}{c}{Atom}
%%          & \multicolumn{1}{c}{$s$} & \multicolumn{2}{c}{$p$} & \multicolumn{2}{c}{$d$} & \multicolumn{2}{c}{Atom} &  \\ \hline
\multirow{2}{*}{Compound} & \multirow{2}{*}{Method} & \multicolumn{7}{c}{$1_{st}$ element} & \multicolumn{7}{c}{$2_{nd}$ element} & \multicolumn{2}{c}{Total}\\ \cline{3-9} \cline{10-16} \cline{17-18} 
&        & \multicolumn{1}{c}{$q_s$} & \multicolumn{1}{c}{$q_p$} & \multicolumn{1}{c}{$m_p$} & \multicolumn{1}{c}{$q_d$} & \multicolumn{1}{c}{$m_d$} & \multicolumn{1}{c}{$q_1$} & \multicolumn{1}{c}{$m_1$}
         & \multicolumn{1}{c}{$q_s$} & \multicolumn{1}{c}{$q_p$} & \multicolumn{1}{c}{$m_p$} & \multicolumn{1}{c}{$q_d$} & \multicolumn{1}{c}{$m_d$} & \multicolumn{1}{c}{$q_2$} & \multicolumn{1}{c}{$m_2$} &  \multicolumn{1}{c}{$q$} & \multicolumn{1}{c}{$m$}  \\ \hline
\mw{FePt}&  VASP  & 0.41  &  0.45 & -0.01 &  6.12 &  2.92 &  6.98 &  2.92 &  0.61  &  0.50 & -0.05 &  7.73 &  0.43 &  8.84 &  0.36 & 15.82 &  3.28 \\
         &  TB    & 0.83  &  0.81 & -0.04 &  6.38 &  2.95 &  8.03 &  2.95 &  0.98  &  0.81 & -0.09 &  8.18 &  0.45 &  9.97 &  0.34 & 18.00 &  3.29 \\ \\ [-1em]
\mw{CoPt}&  VASP  & 0.41  &  0.44 & -0.01 &  7.20 &  1.91 &  8.05 &  1.89 &  0.61  &  0.50 & -0.04 &  7.72 &  0.46 &  8.82 &  0.41 & 16.88 &  2.30 \\
         &  TB    & 0.83  &  0.80 & -0.04 &  7.48 &  1.92 &  9.11 &  1.87 &  0.96  &  0.78 & -0.06 &  8.14 &  0.47 &  9.89 &  0.40 & 19.00 &  2.26 \\ \\ [-1em]
\mw{FePd}&  VASP  & 0.42  &  0.41 & -0.01 &  6.09 &  2.97 &  6.92 &  2.96 &  0.37  &  0.31 & -0.04 &  7.96 &  0.42 &  8.63 &  0.36 & 15.55 &  3.33 \\
         &  TB    & 0.81  &  0.67 & -0.04 &  6.35 &  3.05 &  7.83 &  3.00 &  0.85  &  0.70 & -0.09 &  8.63 &  0.42 & 10.17 &  0.29 & 18.00 &  3.29 \\ \\ [-1em]
\mw{FeNi}&  VASP  & 0.47  &  0.49 & -0.02 &  6.17 &  2.69 &  7.12 &  2.67 &  0.51  &  0.51 & -0.07 &  8.31 &  0.72 &  9.33 &  0.62 & 16.45 &  3.29 \\
         &  TB    & 0.80  &  0.70 & -0.05 &  6.38 &  2.74 &  7.89 &  2.67 &  0.85  &  0.73 & -0.09 &  8.53 &  0.71 & 10.11 &  0.57 & 18.00 &  3.24 \\ \\ [-1em]
\mw{MnGa}&  VASP  & 0.29  &  0.31 &  0.02 &  5.13 &  2.49 &  5.73 &  2.52 &  1.08  &  1.31 & -0.11 &  0.16 &  0.02 &  2.56 & -0.14 &  8.29 &  2.39 \\ 
         &  TB    & 0.55  &  0.47 &  0.02 &  5.45 &  2.60 &  6.46 &  2.64 &  1.36  &  1.89 & -0.15 &  0.28 &  0.03 &  3.54 & -0.15 & 10.00 &  2.49 \\ \\ [-1em]
\mw{MnAl}&  VASP  & 0.33  &  0.34 &  0.02 &  5.12 &  2.37 &  5.78 &  2.41 &  0.40  &  0.42 & -0.04 &  0.00 &  0.00 &  0.82 & -0.06 &  6.60 &  2.35 \\ 
         &  TB    & 0.56  &  0.47 &  0.03 &  5.49 &  2.49 &  6.52 &  2.55 &  1.21  &  1.87 & -0.14 &  0.41 &  0.03 &  3.48 & -0.17 & 10.00 &  2.39 \\
    \hline\hline
\end{tabular*}
\egroup
\end{table*}
\end{widetext}

\bibliography{../../../../../www/refs/bib/references_smpl.bib}

\begin{thebibliography}{50}%
\makeatletter
\providecommand \@ifxundefined [1]{%
 \@ifx{#1\undefined}
}%
\providecommand \@ifnum [1]{%
 \ifnum #1\expandafter \@firstoftwo
 \else \expandafter \@secondoftwo
 \fi
}%
\providecommand \@ifx [1]{%
 \ifx #1\expandafter \@firstoftwo
 \else \expandafter \@secondoftwo
 \fi
}%
\providecommand \natexlab [1]{#1}%
\providecommand \enquote  [1]{``#1''}%
\providecommand \bibnamefont  [1]{#1}%
\providecommand \bibfnamefont [1]{#1}%
\providecommand \citenamefont [1]{#1}%
\providecommand \href@noop [0]{\@secondoftwo}%
\providecommand \href [0]{\begingroup \@sanitize@url \@href}%
\providecommand \@href[1]{\@@startlink{#1}\@@href}%
\providecommand \@@href[1]{\endgroup#1\@@endlink}%
\providecommand \@sanitize@url [0]{\catcode `\\12\catcode `\$12\catcode
  `\&12\catcode `\#12\catcode `\^12\catcode `\_12\catcode `\%12\relax}%
\providecommand \@@startlink[1]{}%
\providecommand \@@endlink[0]{}%
\providecommand \url  [0]{\begingroup\@sanitize@url \@url }%
\providecommand \@url [1]{\endgroup\@href {#1}{\urlprefix }}%
\providecommand \urlprefix  [0]{URL }%
\providecommand \Eprint [0]{\href }%
\providecommand \doibase [0]{http://dx.doi.org/}%
\providecommand \selectlanguage [0]{\@gobble}%
\providecommand \bibinfo  [0]{\@secondoftwo}%
\providecommand \bibfield  [0]{\@secondoftwo}%
\providecommand \translation [1]{[#1]}%
\providecommand \BibitemOpen [0]{}%
\providecommand \bibitemStop [0]{}%
\providecommand \bibitemNoStop [0]{.\EOS\space}%
\providecommand \EOS [0]{\spacefactor3000\relax}%
\providecommand \BibitemShut  [1]{\csname bibitem#1\endcsname}%
\let\auto@bib@innerbib\@empty
%</preamble>
\bibitem [{\citenamefont {{van Vleck}}(1937)}]{van-vleck1937pr}%
  \BibitemOpen
  \bibfield  {author} {\bibinfo {author} {\bibfnamefont {J.~H.}\ \bibnamefont
  {{van Vleck}}},\ }\href@noop {} {\bibfield  {journal} {\bibinfo  {journal}
  {Phys. Rev.}\ }\textbf {\bibinfo {volume} {52}},\ \bibinfo {pages} {1178}
  (\bibinfo {year} {1937})}\BibitemShut {NoStop}%
\bibitem [{\citenamefont {McCallum}\ \emph {et~al.}(2014)\citenamefont
  {McCallum}, \citenamefont {Lewis}, \citenamefont {Skomski}, \citenamefont
  {Kramer},\ and\ \citenamefont {Anderson}}]{mccallum2014armr}%
  \BibitemOpen
  \bibfield  {author} {\bibinfo {author} {\bibfnamefont {R.}~\bibnamefont
  {McCallum}}, \bibinfo {author} {\bibfnamefont {L.}~\bibnamefont {Lewis}},
  \bibinfo {author} {\bibfnamefont {R.}~\bibnamefont {Skomski}}, \bibinfo
  {author} {\bibfnamefont {M.}~\bibnamefont {Kramer}}, \ and\ \bibinfo {author}
  {\bibfnamefont {I.}~\bibnamefont {Anderson}},\ }\href@noop {} {\bibfield
  {journal} {\bibinfo  {journal} {Annual Review of Materials Research}\
  }\textbf {\bibinfo {volume} {44}},\ \bibinfo {pages} {451} (\bibinfo {year}
  {2014})}\BibitemShut {NoStop}%
\bibitem [{\citenamefont {Rau}\ \emph {et~al.}(2014)\citenamefont {Rau},
  \citenamefont {Baumann}, \citenamefont {Rusponi}, \citenamefont {Donati},
  \citenamefont {Stepanow}, \citenamefont {Gragnaniello}, \citenamefont
  {Dreiser}, \citenamefont {Piamonteze}, \citenamefont {Nolting}, \citenamefont
  {Gangopadhyay}, \citenamefont {Albertini}, \citenamefont {Macfarlane},
  \citenamefont {Lutz}, \citenamefont {Jones}, \citenamefont {Gambardella},
  \citenamefont {Heinrich},\ and\ \citenamefont {Brune}}]{rau2014s}%
  \BibitemOpen
  \bibfield  {author} {\bibinfo {author} {\bibfnamefont {I.~G.}\ \bibnamefont
  {Rau}}, \bibinfo {author} {\bibfnamefont {S.}~\bibnamefont {Baumann}},
  \bibinfo {author} {\bibfnamefont {S.}~\bibnamefont {Rusponi}}, \bibinfo
  {author} {\bibfnamefont {F.}~\bibnamefont {Donati}}, \bibinfo {author}
  {\bibfnamefont {S.}~\bibnamefont {Stepanow}}, \bibinfo {author}
  {\bibfnamefont {L.}~\bibnamefont {Gragnaniello}}, \bibinfo {author}
  {\bibfnamefont {J.}~\bibnamefont {Dreiser}}, \bibinfo {author} {\bibfnamefont
  {C.}~\bibnamefont {Piamonteze}}, \bibinfo {author} {\bibfnamefont
  {F.}~\bibnamefont {Nolting}}, \bibinfo {author} {\bibfnamefont
  {S.}~\bibnamefont {Gangopadhyay}}, \bibinfo {author} {\bibfnamefont {O.~R.}\
  \bibnamefont {Albertini}}, \bibinfo {author} {\bibfnamefont {R.~M.}\
  \bibnamefont {Macfarlane}}, \bibinfo {author} {\bibfnamefont {C.~P.}\
  \bibnamefont {Lutz}}, \bibinfo {author} {\bibfnamefont {B.~A.}\ \bibnamefont
  {Jones}}, \bibinfo {author} {\bibfnamefont {P.}~\bibnamefont {Gambardella}},
  \bibinfo {author} {\bibfnamefont {A.~J.}\ \bibnamefont {Heinrich}}, \ and\
  \bibinfo {author} {\bibfnamefont {H.}~\bibnamefont {Brune}},\ }\href@noop {}
  {\bibfield  {journal} {\bibinfo  {journal} {Science}\ }\textbf {\bibinfo
  {volume} {344}},\ \bibinfo {pages} {988} (\bibinfo {year}
  {2014})}\BibitemShut {NoStop}%
\bibitem [{\citenamefont {Koelling}\ and\ \citenamefont
  {Harmon}(1977)}]{koelling1977jpcs}%
  \BibitemOpen
  \bibfield  {author} {\bibinfo {author} {\bibfnamefont {D.~D.}\ \bibnamefont
  {Koelling}}\ and\ \bibinfo {author} {\bibfnamefont {B.~N.}\ \bibnamefont
  {Harmon}},\ }\href@noop {} {\bibfield  {journal} {\bibinfo  {journal}
  {Journal of Physics C: Solid State Physics}\ }\textbf {\bibinfo {volume}
  {10}},\ \bibinfo {pages} {3107} (\bibinfo {year} {1977})}\BibitemShut
  {NoStop}%
\bibitem [{\citenamefont {Li}\ \emph {et~al.}(1990)\citenamefont {Li},
  \citenamefont {Freeman}, \citenamefont {Jansen},\ and\ \citenamefont
  {Fu}}]{li1990prb}%
  \BibitemOpen
  \bibfield  {author} {\bibinfo {author} {\bibfnamefont {C.}~\bibnamefont
  {Li}}, \bibinfo {author} {\bibfnamefont {A.~J.}\ \bibnamefont {Freeman}},
  \bibinfo {author} {\bibfnamefont {H.~J.~F.}\ \bibnamefont {Jansen}}, \ and\
  \bibinfo {author} {\bibfnamefont {C.~L.}\ \bibnamefont {Fu}},\ }\href@noop {}
  {\bibfield  {journal} {\bibinfo  {journal} {Phys. Rev. B}\ }\textbf {\bibinfo
  {volume} {42}},\ \bibinfo {pages} {5433} (\bibinfo {year}
  {1990})}\BibitemShut {NoStop}%
\bibitem [{\citenamefont {Shick}\ \emph {et~al.}(1997)\citenamefont {Shick},
  \citenamefont {Novikov},\ and\ \citenamefont {Freeman}}]{shick1997prb}%
  \BibitemOpen
  \bibfield  {author} {\bibinfo {author} {\bibfnamefont {A.~B.}\ \bibnamefont
  {Shick}}, \bibinfo {author} {\bibfnamefont {D.~L.}\ \bibnamefont {Novikov}},
  \ and\ \bibinfo {author} {\bibfnamefont {A.~J.}\ \bibnamefont {Freeman}},\
  }\href@noop {} {\bibfield  {journal} {\bibinfo  {journal} {Phys. Rev. B}\
  }\textbf {\bibinfo {volume} {56}},\ \bibinfo {pages} {R14259} (\bibinfo
  {year} {1997})}\BibitemShut {NoStop}%
\bibitem [{\citenamefont {Wang}\ \emph
  {et~al.}(1996{\natexlab{a}})\citenamefont {Wang}, \citenamefont {sheng Wang},
  \citenamefont {Wu},\ and\ \citenamefont {Freeman}}]{wang1996jmmm}%
  \BibitemOpen
  \bibfield  {author} {\bibinfo {author} {\bibfnamefont {X.}~\bibnamefont
  {Wang}}, \bibinfo {author} {\bibfnamefont {D.}~\bibnamefont {sheng Wang}},
  \bibinfo {author} {\bibfnamefont {R.}~\bibnamefont {Wu}}, \ and\ \bibinfo
  {author} {\bibfnamefont {A.}~\bibnamefont {Freeman}},\ }\href@noop {}
  {\bibfield  {journal} {\bibinfo  {journal} {Journal of Magnetism and Magnetic
  Materials}\ }\textbf {\bibinfo {volume} {159}},\ \bibinfo {pages} {337 }
  (\bibinfo {year} {1996}{\natexlab{a}})}\BibitemShut {NoStop}%
\bibitem [{\citenamefont {Mackintosh}\ and\ \citenamefont
  {Andersen}(1980)}]{mackintosh1980book}%
  \BibitemOpen
  \bibfield  {author} {\bibinfo {author} {\bibfnamefont {A.}~\bibnamefont
  {Mackintosh}}\ and\ \bibinfo {author} {\bibfnamefont {O.}~\bibnamefont
  {Andersen}},\ }\href@noop {} {\emph {\bibinfo {title} {Electrons at the Fermi
  Surface}}}\ (\bibinfo  {publisher} {Cambridge University Press},\ \bibinfo
  {address} {Cambridge, England},\ \bibinfo {year} {1980})\BibitemShut
  {NoStop}%
\bibitem [{\citenamefont {Weinert}\ \emph {et~al.}(1985)\citenamefont
  {Weinert}, \citenamefont {Watson},\ and\ \citenamefont
  {Davenport}}]{weinert1985prb}%
  \BibitemOpen
  \bibfield  {author} {\bibinfo {author} {\bibfnamefont {M.}~\bibnamefont
  {Weinert}}, \bibinfo {author} {\bibfnamefont {R.~E.}\ \bibnamefont {Watson}},
  \ and\ \bibinfo {author} {\bibfnamefont {J.~W.}\ \bibnamefont {Davenport}},\
  }\href@noop {} {\bibfield  {journal} {\bibinfo  {journal} {Phys. Rev. B}\
  }\textbf {\bibinfo {volume} {32}},\ \bibinfo {pages} {2115} (\bibinfo {year}
  {1985})}\BibitemShut {NoStop}%
\bibitem [{\citenamefont {Daalderop}\ \emph {et~al.}(1990)\citenamefont
  {Daalderop}, \citenamefont {Kelly},\ and\ \citenamefont
  {Schuurmans}}]{daalderop1990prb}%
  \BibitemOpen
  \bibfield  {author} {\bibinfo {author} {\bibfnamefont {G.~H.~O.}\
  \bibnamefont {Daalderop}}, \bibinfo {author} {\bibfnamefont {P.~J.}\
  \bibnamefont {Kelly}}, \ and\ \bibinfo {author} {\bibfnamefont {M.~F.~H.}\
  \bibnamefont {Schuurmans}},\ }\href@noop {} {\bibfield  {journal} {\bibinfo
  {journal} {Phys. Rev. B}\ }\textbf {\bibinfo {volume} {41}},\ \bibinfo
  {pages} {11919} (\bibinfo {year} {1990})}\BibitemShut {NoStop}%
\bibitem [{\citenamefont {Yosida}\ \emph {et~al.}(1965)\citenamefont {Yosida},
  \citenamefont {Okiji},\ and\ \citenamefont {Chikazumi}}]{yosida1965ptp}%
  \BibitemOpen
  \bibfield  {author} {\bibinfo {author} {\bibfnamefont {K.}~\bibnamefont
  {Yosida}}, \bibinfo {author} {\bibfnamefont {A.}~\bibnamefont {Okiji}}, \
  and\ \bibinfo {author} {\bibfnamefont {S.}~\bibnamefont {Chikazumi}},\
  }\href@noop {} {\bibfield  {journal} {\bibinfo  {journal} {Progress of
  Theoretical Physics}\ }\textbf {\bibinfo {volume} {33}},\ \bibinfo {pages}
  {559} (\bibinfo {year} {1965})}\BibitemShut {NoStop}%
\bibitem [{\citenamefont {Abate}\ and\ \citenamefont
  {Asdente}(1965)}]{abate1965pr}%
  \BibitemOpen
  \bibfield  {author} {\bibinfo {author} {\bibfnamefont {E.}~\bibnamefont
  {Abate}}\ and\ \bibinfo {author} {\bibfnamefont {M.}~\bibnamefont
  {Asdente}},\ }\href@noop {} {\bibfield  {journal} {\bibinfo  {journal} {Phys.
  Rev.}\ }\textbf {\bibinfo {volume} {140}},\ \bibinfo {pages} {A1303}
  (\bibinfo {year} {1965})}\BibitemShut {NoStop}%
\bibitem [{\citenamefont {Takayama}\ \emph {et~al.}(1976)\citenamefont
  {Takayama}, \citenamefont {Bohnen},\ and\ \citenamefont
  {Fulde}}]{takayama1976prb}%
  \BibitemOpen
  \bibfield  {author} {\bibinfo {author} {\bibfnamefont {H.}~\bibnamefont
  {Takayama}}, \bibinfo {author} {\bibfnamefont {K.-P.}\ \bibnamefont
  {Bohnen}}, \ and\ \bibinfo {author} {\bibfnamefont {P.}~\bibnamefont
  {Fulde}},\ }\href@noop {} {\bibfield  {journal} {\bibinfo  {journal} {Phys.
  Rev. B}\ }\textbf {\bibinfo {volume} {14}},\ \bibinfo {pages} {2287}
  (\bibinfo {year} {1976})}\BibitemShut {NoStop}%
\bibitem [{\citenamefont {Bruno}(1989)}]{bruno1989prb}%
  \BibitemOpen
  \bibfield  {author} {\bibinfo {author} {\bibfnamefont {P.}~\bibnamefont
  {Bruno}},\ }\href@noop {} {\bibfield  {journal} {\bibinfo  {journal} {Phys.
  Rev. B}\ }\textbf {\bibinfo {volume} {39}},\ \bibinfo {pages} {865} (\bibinfo
  {year} {1989})}\BibitemShut {NoStop}%
\bibitem [{\citenamefont {Cinal}\ \emph {et~al.}(1994)\citenamefont {Cinal},
  \citenamefont {Edwards},\ and\ \citenamefont {Mathon}}]{cinal1994prb}%
  \BibitemOpen
  \bibfield  {author} {\bibinfo {author} {\bibfnamefont {M.}~\bibnamefont
  {Cinal}}, \bibinfo {author} {\bibfnamefont {D.~M.}\ \bibnamefont {Edwards}},
  \ and\ \bibinfo {author} {\bibfnamefont {J.}~\bibnamefont {Mathon}},\
  }\href@noop {} {\bibfield  {journal} {\bibinfo  {journal} {Phys. Rev. B}\
  }\textbf {\bibinfo {volume} {50}},\ \bibinfo {pages} {3754} (\bibinfo {year}
  {1994})}\BibitemShut {NoStop}%
\bibitem [{\citenamefont {{van der Laan}}(1998)}]{van-der-laan1998jpcm}%
  \BibitemOpen
  \bibfield  {author} {\bibinfo {author} {\bibfnamefont {G.}~\bibnamefont {{van
  der Laan}}},\ }\href@noop {} {\bibfield  {journal} {\bibinfo  {journal}
  {Journal of Physics: Condensed Matter}\ }\textbf {\bibinfo {volume} {10}},\
  \bibinfo {pages} {3239} (\bibinfo {year} {1998})}\BibitemShut {NoStop}%
\bibitem [{\citenamefont {Ke}\ and\ \citenamefont {van
  Schilfgaarde}(2015)}]{ke2015prb}%
  \BibitemOpen
  \bibfield  {author} {\bibinfo {author} {\bibfnamefont {L.}~\bibnamefont
  {Ke}}\ and\ \bibinfo {author} {\bibfnamefont {M.}~\bibnamefont {van
  Schilfgaarde}},\ }\href@noop {} {\bibfield  {journal} {\bibinfo  {journal}
  {Phys. Rev. B}\ }\textbf {\bibinfo {volume} {92}},\ \bibinfo {pages} {014423}
  (\bibinfo {year} {2015})}\BibitemShut {NoStop}%
\bibitem [{\citenamefont {Jesche}\ \emph {et~al.}(2015)\citenamefont {Jesche},
  \citenamefont {Ke}, \citenamefont {Jacobs}, \citenamefont {Harmon},
  \citenamefont {Houk},\ and\ \citenamefont {Canfield}}]{jesche2015prb}%
  \BibitemOpen
  \bibfield  {author} {\bibinfo {author} {\bibfnamefont {A.}~\bibnamefont
  {Jesche}}, \bibinfo {author} {\bibfnamefont {L.}~\bibnamefont {Ke}}, \bibinfo
  {author} {\bibfnamefont {J.~L.}\ \bibnamefont {Jacobs}}, \bibinfo {author}
  {\bibfnamefont {B.}~\bibnamefont {Harmon}}, \bibinfo {author} {\bibfnamefont
  {R.~S.}\ \bibnamefont {Houk}}, \ and\ \bibinfo {author} {\bibfnamefont
  {P.~C.}\ \bibnamefont {Canfield}},\ }\href@noop {} {\bibfield  {journal}
  {\bibinfo  {journal} {Phys. Rev. B}\ }\textbf {\bibinfo {volume} {91}},\
  \bibinfo {pages} {180403} (\bibinfo {year} {2015})},\ \bibinfo {note}
  {$\textbf{Rapid communication}$}\BibitemShut {NoStop}%
\bibitem [{\citenamefont {Marzari}\ and\ \citenamefont
  {Vanderbilt}(1997)}]{marzari1997prb}%
  \BibitemOpen
  \bibfield  {author} {\bibinfo {author} {\bibfnamefont {N.}~\bibnamefont
  {Marzari}}\ and\ \bibinfo {author} {\bibfnamefont {D.}~\bibnamefont
  {Vanderbilt}},\ }\href@noop {} {\bibfield  {journal} {\bibinfo  {journal}
  {Phys. Rev. B}\ }\textbf {\bibinfo {volume} {56}},\ \bibinfo {pages} {12847}
  (\bibinfo {year} {1997})}\BibitemShut {NoStop}%
\bibitem [{\citenamefont {Marzari}\ \emph {et~al.}(2012)\citenamefont
  {Marzari}, \citenamefont {Mostofi}, \citenamefont {Yates}, \citenamefont
  {Souza},\ and\ \citenamefont {Vanderbilt}}]{marzari2012rmp}%
  \BibitemOpen
  \bibfield  {author} {\bibinfo {author} {\bibfnamefont {N.}~\bibnamefont
  {Marzari}}, \bibinfo {author} {\bibfnamefont {A.~A.}\ \bibnamefont
  {Mostofi}}, \bibinfo {author} {\bibfnamefont {J.~R.}\ \bibnamefont {Yates}},
  \bibinfo {author} {\bibfnamefont {I.}~\bibnamefont {Souza}}, \ and\ \bibinfo
  {author} {\bibfnamefont {D.}~\bibnamefont {Vanderbilt}},\ }\href@noop {}
  {\bibfield  {journal} {\bibinfo  {journal} {Rev. Mod. Phys.}\ }\textbf
  {\bibinfo {volume} {84}},\ \bibinfo {pages} {1419} (\bibinfo {year}
  {2012})}\BibitemShut {NoStop}%
\bibitem [{\citenamefont {Souza}\ \emph {et~al.}(2001)\citenamefont {Souza},
  \citenamefont {Marzari},\ and\ \citenamefont {Vanderbilt}}]{souza2001prb}%
  \BibitemOpen
  \bibfield  {author} {\bibinfo {author} {\bibfnamefont {I.}~\bibnamefont
  {Souza}}, \bibinfo {author} {\bibfnamefont {N.}~\bibnamefont {Marzari}}, \
  and\ \bibinfo {author} {\bibfnamefont {D.}~\bibnamefont {Vanderbilt}},\
  }\href@noop {} {\bibfield  {journal} {\bibinfo  {journal} {Phys. Rev. B}\
  }\textbf {\bibinfo {volume} {65}},\ \bibinfo {pages} {035109} (\bibinfo
  {year} {2001})}\BibitemShut {NoStop}%
\bibitem [{\citenamefont {Antropov}\ \emph {et~al.}(2014)\citenamefont
  {Antropov}, \citenamefont {Ke},\ and\ \citenamefont
  {\r{A}berg}}]{antropov2014ssc}%
  \BibitemOpen
  \bibfield  {author} {\bibinfo {author} {\bibfnamefont {V.}~\bibnamefont
  {Antropov}}, \bibinfo {author} {\bibfnamefont {L.}~\bibnamefont {Ke}}, \ and\
  \bibinfo {author} {\bibfnamefont {D.}~\bibnamefont {\r{A}berg}},\ }\href@noop
  {} {\bibfield  {journal} {\bibinfo  {journal} {Solid State Communications}\
  }\textbf {\bibinfo {volume} {194}},\ \bibinfo {pages} {35} (\bibinfo {year}
  {2014})}\BibitemShut {NoStop}%
\bibitem [{\citenamefont {Wang}\ \emph
  {et~al.}(1996{\natexlab{b}})\citenamefont {Wang}, \citenamefont {Wu},
  \citenamefont {Wang},\ and\ \citenamefont {Freeman}}]{wang1996prb}%
  \BibitemOpen
  \bibfield  {author} {\bibinfo {author} {\bibfnamefont {X.}~\bibnamefont
  {Wang}}, \bibinfo {author} {\bibfnamefont {R.}~\bibnamefont {Wu}}, \bibinfo
  {author} {\bibfnamefont {D.-s.}\ \bibnamefont {Wang}}, \ and\ \bibinfo
  {author} {\bibfnamefont {A.~J.}\ \bibnamefont {Freeman}},\ }\href@noop {}
  {\bibfield  {journal} {\bibinfo  {journal} {Phys. Rev. B}\ }\textbf {\bibinfo
  {volume} {54}},\ \bibinfo {pages} {61} (\bibinfo {year}
  {1996}{\natexlab{b}})}\BibitemShut {NoStop}%
\bibitem [{\citenamefont {\r{A}berg}\ \emph {et~al.}(2015)\citenamefont
  {\r{A}berg}, \citenamefont {Sadigh},\ and\ \citenamefont
  {Benedict}}]{aberg2015tech-osma}%
  \BibitemOpen
  \bibfield  {author} {\bibinfo {author} {\bibfnamefont {D.}~\bibnamefont
  {\r{A}berg}}, \bibinfo {author} {\bibfnamefont {B.}~\bibnamefont {Sadigh}}, \
  and\ \bibinfo {author} {\bibfnamefont {L.~X.}\ \bibnamefont {Benedict}},\
  }\href@noop {} {\emph {\bibinfo {title} {On the site-decomposition of
  magnetocrystalline anisotropy energy using one-electron eigenstates}}},\
  \bibinfo {type} {Tech. Rep.}\ (\bibinfo  {institution} {Lawrence Livermore
  National Laboratory},\ \bibinfo {year} {2015})\BibitemShut {NoStop}%
\bibitem [{\citenamefont {Ke}\ \emph {et~al.}(2016)\citenamefont {Ke},
  \citenamefont {Kukusta},\ and\ \citenamefont {Johnson}}]{ke2016prbA}%
  \BibitemOpen
  \bibfield  {author} {\bibinfo {author} {\bibfnamefont {L.}~\bibnamefont
  {Ke}}, \bibinfo {author} {\bibfnamefont {D.~A.}\ \bibnamefont {Kukusta}}, \
  and\ \bibinfo {author} {\bibfnamefont {D.~D.}\ \bibnamefont {Johnson}},\
  }\href@noop {} {\bibfield  {journal} {\bibinfo  {journal} {Phys. Rev. B}\
  }\textbf {\bibinfo {volume} {94}},\ \bibinfo {pages} {144429} (\bibinfo
  {year} {2016})}\BibitemShut {NoStop}%
\bibitem [{\citenamefont {Mryasov}\ \emph {et~al.}(2005)\citenamefont
  {Mryasov}, \citenamefont {Nowak}, \citenamefont {Guslienko},\ and\
  \citenamefont {Chantrell}}]{mryasov2005eel}%
  \BibitemOpen
  \bibfield  {author} {\bibinfo {author} {\bibfnamefont {O.~N.}\ \bibnamefont
  {Mryasov}}, \bibinfo {author} {\bibfnamefont {U.}~\bibnamefont {Nowak}},
  \bibinfo {author} {\bibfnamefont {K.~Y.}\ \bibnamefont {Guslienko}}, \ and\
  \bibinfo {author} {\bibfnamefont {R.~W.}\ \bibnamefont {Chantrell}},\
  }\href@noop {} {\bibfield  {journal} {\bibinfo  {journal} {EPL (Europhysics
  Letters)}\ }\textbf {\bibinfo {volume} {69}},\ \bibinfo {pages} {805}
  (\bibinfo {year} {2005})}\BibitemShut {NoStop}%
\bibitem [{\citenamefont {Daalderop}\ \emph {et~al.}(1991)\citenamefont
  {Daalderop}, \citenamefont {Kelly},\ and\ \citenamefont
  {Schuurmans}}]{daalderop1991prb}%
  \BibitemOpen
  \bibfield  {author} {\bibinfo {author} {\bibfnamefont {G.~H.~O.}\
  \bibnamefont {Daalderop}}, \bibinfo {author} {\bibfnamefont {P.~J.}\
  \bibnamefont {Kelly}}, \ and\ \bibinfo {author} {\bibfnamefont {M.~F.~H.}\
  \bibnamefont {Schuurmans}},\ }\href@noop {} {\bibfield  {journal} {\bibinfo
  {journal} {Phys. Rev. B}\ }\textbf {\bibinfo {volume} {44}},\ \bibinfo
  {pages} {12054} (\bibinfo {year} {1991})}\BibitemShut {NoStop}%
\bibitem [{\citenamefont {Solovyev}\ \emph {et~al.}(1995)\citenamefont
  {Solovyev}, \citenamefont {Dederichs},\ and\ \citenamefont
  {Mertig}}]{solovyev1995prb}%
  \BibitemOpen
  \bibfield  {author} {\bibinfo {author} {\bibfnamefont {I.~V.}\ \bibnamefont
  {Solovyev}}, \bibinfo {author} {\bibfnamefont {P.~H.}\ \bibnamefont
  {Dederichs}}, \ and\ \bibinfo {author} {\bibfnamefont {I.}~\bibnamefont
  {Mertig}},\ }\href@noop {} {\bibfield  {journal} {\bibinfo  {journal} {Phys.
  Rev. B}\ }\textbf {\bibinfo {volume} {52}},\ \bibinfo {pages} {13419}
  (\bibinfo {year} {1995})}\BibitemShut {NoStop}%
\bibitem [{\citenamefont {Kota}\ and\ \citenamefont
  {Sakuma}(2014)}]{kota2014jpsj}%
  \BibitemOpen
  \bibfield  {author} {\bibinfo {author} {\bibfnamefont {Y.}~\bibnamefont
  {Kota}}\ and\ \bibinfo {author} {\bibfnamefont {A.}~\bibnamefont {Sakuma}},\
  }\href@noop {} {\bibfield  {journal} {\bibinfo  {journal} {Journal of the
  Physical Society of Japan}\ }\textbf {\bibinfo {volume} {83}},\ \bibinfo
  {pages} {034715} (\bibinfo {year} {2014})}\BibitemShut {NoStop}%
\bibitem [{\citenamefont {Kresse}\ and\ \citenamefont
  {Hafner}(1993)}]{kresse1993prb}%
  \BibitemOpen
  \bibfield  {author} {\bibinfo {author} {\bibfnamefont {G.}~\bibnamefont
  {Kresse}}\ and\ \bibinfo {author} {\bibfnamefont {J.}~\bibnamefont
  {Hafner}},\ }\href@noop {} {\bibfield  {journal} {\bibinfo  {journal} {Phys.
  Rev. B}\ }\textbf {\bibinfo {volume} {47}},\ \bibinfo {pages} {558} (\bibinfo
  {year} {1993})}\BibitemShut {NoStop}%
\bibitem [{\citenamefont {Kresse}\ and\ \citenamefont
  {Furthm\"uller}(1996)}]{kresse1996prb}%
  \BibitemOpen
  \bibfield  {author} {\bibinfo {author} {\bibfnamefont {G.}~\bibnamefont
  {Kresse}}\ and\ \bibinfo {author} {\bibfnamefont {J.}~\bibnamefont
  {Furthm\"uller}},\ }\href@noop {} {\bibfield  {journal} {\bibinfo  {journal}
  {Phys. Rev. B}\ }\textbf {\bibinfo {volume} {54}},\ \bibinfo {pages} {11169}
  (\bibinfo {year} {1996})}\BibitemShut {NoStop}%
\bibitem [{\citenamefont {Methfessel}\ \emph {et~al.}(2000)\citenamefont
  {Methfessel}, \citenamefont {van Schilfgaarde},\ and\ \citenamefont
  {Casali}}]{methfessel2000coll-c3fl}%
  \BibitemOpen
  \bibfield  {author} {\bibinfo {author} {\bibfnamefont {M.}~\bibnamefont
  {Methfessel}}, \bibinfo {author} {\bibfnamefont {M.}~\bibnamefont {van
  Schilfgaarde}}, \ and\ \bibinfo {author} {\bibfnamefont {R.~A.}\ \bibnamefont
  {Casali}},\ }in\ \href@noop {} {\emph {\bibinfo {booktitle} {Electronic
  Structure and Physical Properties of Solids: The Uses of the LMTO Method}}},\
  \bibinfo {series} {Lecture Notes in Physics}, Vol.\ \bibinfo {volume} {535},\
  \bibinfo {editor} {edited by\ \bibinfo {editor} {\bibfnamefont
  {H.}~\bibnamefont {Dreysse}}}\ (\bibinfo  {publisher} {Springer-Verlag,
  Berlin},\ \bibinfo {year} {2000})\BibitemShut {NoStop}%
\bibitem [{\citenamefont {Perdew}\ \emph {et~al.}(1996)\citenamefont {Perdew},
  \citenamefont {Burke},\ and\ \citenamefont {Ernzerhof}}]{perdew1996prl}%
  \BibitemOpen
  \bibfield  {author} {\bibinfo {author} {\bibfnamefont {J.~P.}\ \bibnamefont
  {Perdew}}, \bibinfo {author} {\bibfnamefont {K.}~\bibnamefont {Burke}}, \
  and\ \bibinfo {author} {\bibfnamefont {M.}~\bibnamefont {Ernzerhof}},\
  }\href@noop {} {\bibfield  {journal} {\bibinfo  {journal} {Phys. Rev. Lett.}\
  }\textbf {\bibinfo {volume} {77}},\ \bibinfo {pages} {3865} (\bibinfo {year}
  {1996})}\BibitemShut {NoStop}%
\bibitem [{\citenamefont {{von Barth}}\ and\ \citenamefont
  {Hedin}(1972)}]{von-barth1972jpcs}%
  \BibitemOpen
  \bibfield  {author} {\bibinfo {author} {\bibfnamefont {U.}~\bibnamefont {{von
  Barth}}}\ and\ \bibinfo {author} {\bibfnamefont {L.}~\bibnamefont {Hedin}},\
  }\href@noop {} {\bibfield  {journal} {\bibinfo  {journal} {Journal of Physics
  C: Solid State Physics}\ }\textbf {\bibinfo {volume} {5}},\ \bibinfo {pages}
  {1629} (\bibinfo {year} {1972})}\BibitemShut {NoStop}%
\bibitem [{\citenamefont {Mostofi}\ \emph {et~al.}(2014)\citenamefont
  {Mostofi}, \citenamefont {Yates}, \citenamefont {Pizzi}, \citenamefont {Lee},
  \citenamefont {Souza}, \citenamefont {Vanderbilt},\ and\ \citenamefont
  {Marzari}}]{mostofi2014cpc}%
  \BibitemOpen
  \bibfield  {author} {\bibinfo {author} {\bibfnamefont {A.~A.}\ \bibnamefont
  {Mostofi}}, \bibinfo {author} {\bibfnamefont {J.~R.}\ \bibnamefont {Yates}},
  \bibinfo {author} {\bibfnamefont {G.}~\bibnamefont {Pizzi}}, \bibinfo
  {author} {\bibfnamefont {Y.-S.}\ \bibnamefont {Lee}}, \bibinfo {author}
  {\bibfnamefont {I.}~\bibnamefont {Souza}}, \bibinfo {author} {\bibfnamefont
  {D.}~\bibnamefont {Vanderbilt}}, \ and\ \bibinfo {author} {\bibfnamefont
  {N.}~\bibnamefont {Marzari}},\ }\href@noop {} {\bibfield  {journal} {\bibinfo
   {journal} {Computer Physics Communications}\ }\textbf {\bibinfo {volume}
  {185}},\ \bibinfo {pages} {2309} (\bibinfo {year} {2014})}\BibitemShut
  {NoStop}%
\bibitem [{\citenamefont {Bl\"ochl}\ \emph {et~al.}(1994)\citenamefont
  {Bl\"ochl}, \citenamefont {Jepsen},\ and\ \citenamefont
  {Andersen}}]{blochl1994prb}%
  \BibitemOpen
  \bibfield  {author} {\bibinfo {author} {\bibfnamefont {P.~E.}\ \bibnamefont
  {Bl\"ochl}}, \bibinfo {author} {\bibfnamefont {O.}~\bibnamefont {Jepsen}}, \
  and\ \bibinfo {author} {\bibfnamefont {O.~K.}\ \bibnamefont {Andersen}},\
  }\href@noop {} {\bibfield  {journal} {\bibinfo  {journal} {Phys. Rev. B}\
  }\textbf {\bibinfo {volume} {49}},\ \bibinfo {pages} {16223} (\bibinfo {year}
  {1994})}\BibitemShut {NoStop}%
\bibitem [{\citenamefont {Barreteau}\ \emph {et~al.}(2016)\citenamefont
  {Barreteau}, \citenamefont {Spanjaard},\ and\ \citenamefont
  {Desjonqu\`{e}res}}]{barreteau2016crp}%
  \BibitemOpen
  \bibfield  {author} {\bibinfo {author} {\bibfnamefont {C.}~\bibnamefont
  {Barreteau}}, \bibinfo {author} {\bibfnamefont {D.}~\bibnamefont
  {Spanjaard}}, \ and\ \bibinfo {author} {\bibfnamefont {M.-C.}\ \bibnamefont
  {Desjonqu\`{e}res}},\ }\href@noop {} {\bibfield  {journal} {\bibinfo
  {journal} {Comptes Rendus Physique}\ }\textbf {\bibinfo {volume} {17}},\
  \bibinfo {pages} {406} (\bibinfo {year} {2016})},\ \bibinfo {note} {condensed
  matter physics in the 21st century: The legacy of Jacques
  Friedel}\BibitemShut {NoStop}%
\bibitem [{\citenamefont {Subkow}\ and\ \citenamefont
  {F\"ahnle}(2009)}]{subkow2009prb}%
  \BibitemOpen
  \bibfield  {author} {\bibinfo {author} {\bibfnamefont {S.}~\bibnamefont
  {Subkow}}\ and\ \bibinfo {author} {\bibfnamefont {M.}~\bibnamefont
  {F\"ahnle}},\ }\href@noop {} {\bibfield  {journal} {\bibinfo  {journal}
  {Phys. Rev. B}\ }\textbf {\bibinfo {volume} {80}},\ \bibinfo {pages} {212404}
  (\bibinfo {year} {2009})}\BibitemShut {NoStop}%
\bibitem [{\citenamefont {Zeller}\ \emph {et~al.}(1982)\citenamefont {Zeller},
  \citenamefont {Deutz},\ and\ \citenamefont {Dederichs}}]{zeller1982ssc}%
  \BibitemOpen
  \bibfield  {author} {\bibinfo {author} {\bibfnamefont {R.}~\bibnamefont
  {Zeller}}, \bibinfo {author} {\bibfnamefont {J.}~\bibnamefont {Deutz}}, \
  and\ \bibinfo {author} {\bibfnamefont {P.}~\bibnamefont {Dederichs}},\
  }\href@noop {} {\bibfield  {journal} {\bibinfo  {journal} {Solid State
  Communications}\ }\textbf {\bibinfo {volume} {44}},\ \bibinfo {pages} {993 }
  (\bibinfo {year} {1982})}\BibitemShut {NoStop}%
\bibitem [{\citenamefont {Zemen}\ \emph {et~al.}(2014)\citenamefont {Zemen},
  \citenamefont {Mašek}, \citenamefont {Kučera}, \citenamefont {Mol},
  \citenamefont {Motloch},\ and\ \citenamefont {Jungwirth}}]{zemen2014jmmm}%
  \BibitemOpen
  \bibfield  {author} {\bibinfo {author} {\bibfnamefont {J.}~\bibnamefont
  {Zemen}}, \bibinfo {author} {\bibfnamefont {J.}~\bibnamefont {Mašek}},
  \bibinfo {author} {\bibfnamefont {J.}~\bibnamefont {Kučera}}, \bibinfo
  {author} {\bibfnamefont {J.}~\bibnamefont {Mol}}, \bibinfo {author}
  {\bibfnamefont {P.}~\bibnamefont {Motloch}}, \ and\ \bibinfo {author}
  {\bibfnamefont {T.}~\bibnamefont {Jungwirth}},\ }\href@noop {} {\bibfield
  {journal} {\bibinfo  {journal} {Journal of Magnetism and Magnetic Materials}\
  }\textbf {\bibinfo {volume} {356}},\ \bibinfo {pages} {87} (\bibinfo {year}
  {2014})}\BibitemShut {NoStop}%
\bibitem [{\citenamefont {Heine}\ \emph {et~al.}(1984)\citenamefont {Heine},
  \citenamefont {Kok},\ and\ \citenamefont {Nex}}]{heine1984jmmm}%
  \BibitemOpen
  \bibfield  {author} {\bibinfo {author} {\bibfnamefont {V.}~\bibnamefont
  {Heine}}, \bibinfo {author} {\bibfnamefont {W.}~\bibnamefont {Kok}}, \ and\
  \bibinfo {author} {\bibfnamefont {C.}~\bibnamefont {Nex}},\ }\href@noop {}
  {\bibfield  {journal} {\bibinfo  {journal} {Journal of Magnetism and Magnetic
  Materials}\ }\textbf {\bibinfo {volume} {43}},\ \bibinfo {pages} {61}
  (\bibinfo {year} {1984})}\BibitemShut {NoStop}%
\bibitem [{\citenamefont {Inoue}\ \emph {et~al.}(2015)\citenamefont {Inoue},
  \citenamefont {Yoshioka},\ and\ \citenamefont {Tsuchiura}}]{inoue2015jap}%
  \BibitemOpen
  \bibfield  {author} {\bibinfo {author} {\bibfnamefont {J.-I.}\ \bibnamefont
  {Inoue}}, \bibinfo {author} {\bibfnamefont {T.}~\bibnamefont {Yoshioka}}, \
  and\ \bibinfo {author} {\bibfnamefont {H.}~\bibnamefont {Tsuchiura}},\
  }\href@noop {} {\bibfield  {journal} {\bibinfo  {journal} {Journal of Applied
  Physics}\ }\textbf {\bibinfo {volume} {117}},\ \bibinfo {pages} {17C720}
  (\bibinfo {year} {2015})}\BibitemShut {NoStop}%
\bibitem [{\citenamefont {Daalderop}\ \emph {et~al.}(1994)\citenamefont
  {Daalderop}, \citenamefont {Kelly},\ and\ \citenamefont
  {Schuurmans}}]{daalderop1994prb}%
  \BibitemOpen
  \bibfield  {author} {\bibinfo {author} {\bibfnamefont {G.~H.~O.}\
  \bibnamefont {Daalderop}}, \bibinfo {author} {\bibfnamefont {P.~J.}\
  \bibnamefont {Kelly}}, \ and\ \bibinfo {author} {\bibfnamefont {M.~F.~H.}\
  \bibnamefont {Schuurmans}},\ }\href@noop {} {\bibfield  {journal} {\bibinfo
  {journal} {Phys. Rev. B}\ }\textbf {\bibinfo {volume} {50}},\ \bibinfo
  {pages} {9989} (\bibinfo {year} {1994})}\BibitemShut {NoStop}%
\bibitem [{\citenamefont {Wang}\ \emph {et~al.}(1993)\citenamefont {Wang},
  \citenamefont {Wu},\ and\ \citenamefont {Freeman}}]{wang1993prb}%
  \BibitemOpen
  \bibfield  {author} {\bibinfo {author} {\bibfnamefont {D.-s.}\ \bibnamefont
  {Wang}}, \bibinfo {author} {\bibfnamefont {R.}~\bibnamefont {Wu}}, \ and\
  \bibinfo {author} {\bibfnamefont {A.~J.}\ \bibnamefont {Freeman}},\
  }\href@noop {} {\bibfield  {journal} {\bibinfo  {journal} {Phys. Rev. B}\
  }\textbf {\bibinfo {volume} {47}},\ \bibinfo {pages} {14932} (\bibinfo {year}
  {1993})}\BibitemShut {NoStop}%
\bibitem [{\citenamefont {Moos}\ \emph {et~al.}(1996)\citenamefont {Moos},
  \citenamefont {Hübner},\ and\ \citenamefont {Bennemann}}]{moos1996ssc}%
  \BibitemOpen
  \bibfield  {author} {\bibinfo {author} {\bibfnamefont {T.}~\bibnamefont
  {Moos}}, \bibinfo {author} {\bibfnamefont {W.}~\bibnamefont {Hübner}}, \
  and\ \bibinfo {author} {\bibfnamefont {K.}~\bibnamefont {Bennemann}},\
  }\href@noop {} {\bibfield  {journal} {\bibinfo  {journal} {Solid State
  Communications}\ }\textbf {\bibinfo {volume} {98}},\ \bibinfo {pages} {639}
  (\bibinfo {year} {1996})}\BibitemShut {NoStop}%
\bibitem [{\citenamefont {Ravindran}\ \emph {et~al.}(2001)\citenamefont
  {Ravindran}, \citenamefont {Kjekshus}, \citenamefont {Fjellv\aa{}g},
  \citenamefont {James}, \citenamefont {Nordstr\"om}, \citenamefont
  {Johansson},\ and\ \citenamefont {Eriksson}}]{ravindran2001prb}%
  \BibitemOpen
  \bibfield  {author} {\bibinfo {author} {\bibfnamefont {P.}~\bibnamefont
  {Ravindran}}, \bibinfo {author} {\bibfnamefont {A.}~\bibnamefont {Kjekshus}},
  \bibinfo {author} {\bibfnamefont {H.}~\bibnamefont {Fjellv\aa{}g}}, \bibinfo
  {author} {\bibfnamefont {P.}~\bibnamefont {James}}, \bibinfo {author}
  {\bibfnamefont {L.}~\bibnamefont {Nordstr\"om}}, \bibinfo {author}
  {\bibfnamefont {B.}~\bibnamefont {Johansson}}, \ and\ \bibinfo {author}
  {\bibfnamefont {O.}~\bibnamefont {Eriksson}},\ }\href@noop {} {\bibfield
  {journal} {\bibinfo  {journal} {Phys. Rev. B}\ }\textbf {\bibinfo {volume}
  {63}},\ \bibinfo {pages} {144409} (\bibinfo {year} {2001})}\BibitemShut
  {NoStop}%
\bibitem [{\citenamefont {\v{S}ipr}\ \emph {et~al.}(2014)\citenamefont
  {\v{S}ipr}, \citenamefont {Bornemann}, \citenamefont {Ebert},\ and\
  \citenamefont {Min\'{a}r}}]{sipr2014jpcm}%
  \BibitemOpen
  \bibfield  {author} {\bibinfo {author} {\bibfnamefont {O.}~\bibnamefont
  {\v{S}ipr}}, \bibinfo {author} {\bibfnamefont {S.}~\bibnamefont {Bornemann}},
  \bibinfo {author} {\bibfnamefont {H.}~\bibnamefont {Ebert}}, \ and\ \bibinfo
  {author} {\bibfnamefont {J.}~\bibnamefont {Min\'{a}r}},\ }\href@noop {}
  {\bibfield  {journal} {\bibinfo  {journal} {Journal of Physics: Condensed
  Matter}\ }\textbf {\bibinfo {volume} {26}},\ \bibinfo {pages} {196002}
  (\bibinfo {year} {2014})}\BibitemShut {NoStop}%
\bibitem [{\citenamefont {Ayaz~Khan}\ \emph {et~al.}(2016)\citenamefont
  {Ayaz~Khan}, \citenamefont {Blaha}, \citenamefont {Ebert}, \citenamefont
  {Min\'ar},\ and\ \citenamefont {\ifmmode~\check{S}\else
  \v{S}\fi{}ipr}}]{ayaz-khan2016prb}%
  \BibitemOpen
  \bibfield  {author} {\bibinfo {author} {\bibfnamefont {S.}~\bibnamefont
  {Ayaz~Khan}}, \bibinfo {author} {\bibfnamefont {P.}~\bibnamefont {Blaha}},
  \bibinfo {author} {\bibfnamefont {H.}~\bibnamefont {Ebert}}, \bibinfo
  {author} {\bibfnamefont {J.}~\bibnamefont {Min\'ar}}, \ and\ \bibinfo
  {author} {\bibfnamefont {O.~c.~v.}\ \bibnamefont {\ifmmode~\check{S}\else
  \v{S}\fi{}ipr}},\ }\href@noop {} {\bibfield  {journal} {\bibinfo  {journal}
  {Phys. Rev. B}\ }\textbf {\bibinfo {volume} {94}},\ \bibinfo {pages} {144436}
  (\bibinfo {year} {2016})}\BibitemShut {NoStop}%
\bibitem [{\citenamefont {{van Schilfgaarde}}\ \emph
  {et~al.}(2006)\citenamefont {{van Schilfgaarde}}, \citenamefont {Kotani},\
  and\ \citenamefont {Faleev}}]{van-schilfgaarde2006prl}%
  \BibitemOpen
  \bibfield  {author} {\bibinfo {author} {\bibfnamefont {M.}~\bibnamefont {{van
  Schilfgaarde}}}, \bibinfo {author} {\bibfnamefont {T.}~\bibnamefont
  {Kotani}}, \ and\ \bibinfo {author} {\bibfnamefont {S.}~\bibnamefont
  {Faleev}},\ }\href@noop {} {\bibfield  {journal} {\bibinfo  {journal} {Phys.
  Rev. Lett.}\ }\textbf {\bibinfo {volume} {96}},\ \bibinfo {pages} {226402}
  (\bibinfo {year} {2006})}\BibitemShut {NoStop}%
\bibitem [{\citenamefont {Kotani}\ \emph {et~al.}(2007)\citenamefont {Kotani},
  \citenamefont {van Schilfgaarde},\ and\ \citenamefont
  {Faleev}}]{kotani2007prb}%
  \BibitemOpen
  \bibfield  {author} {\bibinfo {author} {\bibfnamefont {T.}~\bibnamefont
  {Kotani}}, \bibinfo {author} {\bibfnamefont {M.}~\bibnamefont {van
  Schilfgaarde}}, \ and\ \bibinfo {author} {\bibfnamefont {S.~V.}\ \bibnamefont
  {Faleev}},\ }\href@noop {} {\bibfield  {journal} {\bibinfo  {journal} {Phys.
  Rev. B}\ }\textbf {\bibinfo {volume} {76}},\ \bibinfo {pages} {165106}
  (\bibinfo {year} {2007})}\BibitemShut {NoStop}%
\end{thebibliography}%
\bigskip

\end{document}